\documentclass[aps,showpacs,showkeys]{revtex4} 
\usepackage{amsfonts}
\usepackage{amsmath}
\usepackage{amssymb}
\usepackage{graphicx}%
\usepackage{tensor} 	
\usepackage{slashed}	
\usepackage{braket,mleftright}
\usepackage{hyperref}

\newcommand{\spinor}{state vector}
\newcommand{\spinors}{state vectors}

\begin{document}
\title{Parity-based formalism for high spin matter fields.}
\author{M. Napsuciale and Selim G\'{o}mez-\'{A}vila}
\affiliation{Departamento de F\'{\i}sica, Universidad de Guanajuato, Lomas del Bosque 103,
Fraccionamiento Lomas del Campestre, 37150, Le\'{o}n, Guanajuato, M\'{e}xico.}

\begin{abstract}
Using the recent parity-based construction of a covariant basis for operators acting on the $(j,0)\oplus(0,j)$ representation of the 
Homogeneous Lorentz Group, we propose a formalism for the description of high spin {\it matter} fields, based on the projection over 
subspaces of well-defined parity. We identify two possibilities for the projection (on-shell and off-shell projection) which in general 
yield equivalent free theories but different interacting theories. For all $j$ except for $j=1/2$, we find that the projection does not 
completely fix the properties of the interacting theory. This freedom is related to the fact that the covariant form of parity can be 
written in terms of one of the symmetric traceless tensors in the covariant basis and in general allows for a free magnetic dipole term 
in the lagrangian. We gauge the theory and construct the charge conjugation operator  showing that it commutes with parity for bosons, 
and anticommute in the case of fermions as expected. In the case of bosons, the parity invariant subspaces are also invariant under 
charge conjugation and time reversal and the formulation of a quantum field theory can be done using only these subspaces. 

As a first exhaustive example we work out the electrodynamics for $j=1$ {\it matter} bosons, rewrite the theory in terms of an 
antisymmetric tensor field and compare our results with existing formalisms in the literature, either in tensor or ``spinor'' language. 
We find that there are three essentially different formalisms: i) formalisms equivalent to the on-shell parity projection, 
ii) formalisms equivalent to the off-shell parity projection and iii) the Poincar\'e projector formalism which describes a degenerate 
parity doublet. In particular, the tensor formalism used in chiral perturbation theory with resonances ($R\chi PT$) is the same as a 
theory proposed by Shay and Good in ``spinor'' language and corresponds to our theory based on the off-shell parity projection. Also, a 
theory proposed by Joos and Weinberg in``spinor'' basis is the same as the ``antisymmetric tensor matter field'' formalism used by 
Chizhov (in the massless case) and corresponds to our on-shell parity projection. 

Naive power counting admits anomalous magnetic-dipole terms and self-interactions at tree level. We perform a chiral decomposition of 
these theories and show that chiral symmetry can be realized linearly only for the theory based on the on-shell projection. Chiral 
symmetry forbids mass and anomalous magnetic dipole terms and in general admits six self-interaction terms.  We conclude that this 
is the appropriate framework to attempt the incorporation of spin 1 {\it matter} bosons in chiral theories like the standard model.
\end{abstract}
\pacs{03.65.Pm, 11.30.Cp, 11.30.Er}
\keywords{High spin, parity}
\maketitle

\section{Introduction.}
The largely awaited first results of the Large Hadron Collider (LHC) for the structure of the Higgs sector have been released, 
confirming the existence of a scalar particle with a mass of 126 GeV, as required by precision data of the standard model 
\cite{Aad20121, Chatrchyan201230}. These results put serious constraints on the existence of supersymmetric particles, large 
extra dimensions and other proposals for physics beyond the standard model \cite{Shifman:2012na}. 

In this context, it is reasonable to explore different possibilities for the existence of new particles. It is remarkable that 
in the standard model there is a clear correspondence of types of fields with representations of the Homogeneous Lorentz Group (HLG). 
Matter fields transform in the $(\frac{1}{2},0)$ and $(0,\frac{1}{2})$ representation, gauge fields in the $(\frac{1}{2},\frac{1}{2})$ 
and the Higgs, the particle that endows the matter (and some of the gauge) fields with mass, transforms in the $(0,0)$ representation; 
only  particles of spin $0$, $\frac{1}{2}$ and $1$ are present. The most appealing extensions of the standard model consider the 
enlargement of the gauge group with the incorporation of the only space-time symmetry beyond the Poincar\'{e} symmetries, namely, 
supersymmetry. Although in the supersymmetric extensions of the standard  model particles with spin $\frac{3}{2}$ appear, they 
enter the formalism as gauge fields transforming in the $(1, \frac{1}{2}) \oplus (\frac{1}{2},1)$ representation of the HLG, not 
as matter fields. 

The tension between the existing proposals for physics beyond the standard model and the recent results of the LHC, together with 
the aforementioned correspondence of the standard model building blocks with specific representations of the Lorentz group, encourages 
us to study the incorporation of high spin \textit{matter} fields in the formulation of theories beyond the standard model. Guided by 
this correspondence, we posit that high spin matter elementary particles, if realized in nature, should be described by fields 
transforming in the $(j,0)\oplus(0,j)$ representation of the HLG. The first step towards the incorporation of these particles in 
extensions of the standard model is to have a consistent description of such fields. 

The problems in the description of high spin particles  \cite{Johnson:1960vt, Velo:1970ur, Velo:1969bt} have been addressed  
by many authors, leading to a  variety of formulations for high spin 
\cite{Joos:1962qq, Weinberg:1964cn, {Weaver:1964zz}, {Tung:1967zz}, {Shay:1968iq}, {Hammer:1968zz}, {Seetharaman:1971nz}, 
{Seetharaman:1971rg}, {Tucker:1971bi}} which have eventually been found unsatisfactory in  general \cite{Eeg:1972us, Eeg:1973xs}. 
The interest on this topic decreased upon the completion of the standard model at the beginning of the seventies, when the focus of 
the community was put on the precision phenomenology of the standard model, in the uncovering of the nature of the non-perturbative 
regime of QCD, and in the formulation of the extension of the ideas behind the standard model to new scenarios which did not require 
to address the problems of the description of high spin particles.

In this work we propose a formalism for the description of high spin matter fields transforming in the $(j,0)\oplus(0,j)$ family of 
representations. Our approach is motivated by an alternative view of the Dirac theory as a manifestation of the discrete symmetries 
satisfied by the interactions of the particle in the representation in which the field transforms. This view has been put forth before 
\cite{Kirchbach:2004qn, DelgadoAcosta:2010nx}, and is in the spirit of the quantum theory of  fields developed by Weinberg\cite{Weinberg:1995mt}. 
To be specific, we take the position that the free theory for spin $j$ and well defined parity is dictated by the properties of the 
representation chosen and by the projection over its parity subspaces.  In order to properly realize charge conjugation and time reversal 
we use as the equation of motion  the projection condition, and we gauge this equation using electromagnetism as a simple example. 
Then we construct charge conjugation showing that, as expected, this operator commutes with parity for bosons and anticommutes for fermions. 

In this construction we actually find two possibilities for the realization of the parity projection, associated to the on-shell projection 
(projection operation valid only in the case when the on-shell relation $p^2=m^2$ holds) and off-shell projection (projection operation valid
for arbitrary values of $p^2$). We find that the on-shell formulation reproduces Dirac theory in the case $j=1/2$ and this is the appropriate
framework for the description of high spin fermion fields. As for bosonic fields both approaches are feasible yielding equivalent free field 
theories but different interacting theories. We also find that parity projection fixes completely the properties of the interacting theory 
only in the case $j=1/2$. For $j\geq 1$ the parity projection used as a dynamical principle yields ambiguities related to the fact that the 
covariant form of parity operator involves only a symmetric space-time tensor. Solving this ambiguity requires to introduce an additional 
anti-symmetric term to which the free theory is insensitive, introducing a free parameter in the theory.

As a first exhaustive example, we work out our formalism for the electrodynamics of spin $1$ matter bosons. We consider both possibilities 
(on-shell and off-shell projection) for the interacting theory, and performing a Gordon-like decomposition of the corresponding electromagnetic 
currents we show that the free parameters are actually ``anomalous'' magnetic moments related to the existence of a Pauli term in the corresponding 
Lagrangian. This term is gauge invariant and due to mass dimension one of the field, turns out to have mass dimension four thus it must be 
included in the naively renormalizable lagrangian anyway. 

We translate out theory to the antisymmetric tensor language and compare our formalism  with existing formulations in the literature for the 
$(1,0)\oplus(0,1)$ representation, either in terms  of tensor or spinor fields. 

Concerning formulations in spinor language, the on-shell 
formulation in the case of positive parity and for a vanishing ``anomalous'' magnetic moment, coincides with a proposal by 
Joos \cite{Joos:1962qq} and Weinberg \cite{Weinberg:1964cn} . The off-shell formulation, with a vanishing ``anomalous'' magnetic moment, 
corresponds to a theory considered previously in \cite{Shay:1968iq} \cite{Tucker:1971bi} in spinor language. This theory with a nonvanishing 
``anomalous'' magnetic moment was analyzed in \cite{Prabhakaran:1973pr}, where it is shown that the classical theory is causal only for a total 
gyromagnetic factor $g=1$. 

As to tensor language, the most general second order lagrangian written down in terms of tensor fields was considered in \cite{Ecker:1988te}. 
There, appropriate constraints were imposed in order to have a theory with a single pole in the propagator, and the formalism was used to 
formulate effective theories for light hadrons in a non-linear implementation of chiral symmetry. We find that this formalism is the tensor 
language version of the theory proposed in \cite{Shay:1968iq} for either positive or negative parity. We show that there is another possible  
theory arising from the most general lagrangian in tensor language written in \cite{Ecker:1988te} and not considered there. This theory 
corresponds with the Poincar\'e  projector formalism for spin 1 matter bosons discussed in \cite{Napsuciale:2006wr,DelgadoAcosta:2012yc}, 
which has been recently shown to describe a parity doublet \cite{Delgado-Acosta:2013nia}. In this last reference the parity projector in 
$(1,0)\oplus (0,1)$ is also introduced in tensor language and it is extracted from an explicit construction of the states  induced as 
derivatives of the states in the $(1/2,1/2)$ representation. This formalism corresponds to an ``anomalous'' gyromagnetic factor of $1/2$ 
(total gyromagnetic factor $g=1$) of our theory based on the off-shell parity projection. In \cite{Delgado-Acosta:2013nia} it has been 
shown that at the classical level this theory is causal in agreement with results in \cite{Prabhakaran:1973pr}.
 
We analyze the chiral properties of our formalism finding that the formulation based on the on-shell projection is the only one 
admitting a linear realization of chiral symmetry in the massless case. Therefore, in the massless limit this formalism is the only one 
suitable for the description of chiral matter spin 1 bosons. 
 
Our paper is organized as follows. In the next section we describe the structure of the $(j,0)\oplus(0,j)$ representation. In section III 
we describe the general structure for the covariant basis for operators acting on this representation. In section IV we construct the 
projectors over subspaces of definite parity and give the Lagrangian for the free field theory. Interactions are introduced in section V, 
where a comprehensive discussion of the discrete symmetries of the theory is given. We work out the electrodynamics of spin 1 matter bosons 
in section VI. In section VII we rewrite our theory in the tensor basis for the $(1,0)\oplus(0,1)$ representation and make a detailed 
comparison with different proposals in the literature both in spinor and tensor language. The chiral structure of the two theories is 
discussed in section VIII and our conclusions are given in section IX.

\section{Structure of the \texorpdfstring{$(j,0)\oplus(0,j)$} representation.}

It is well known that the HLG is locally isomorphic to $SU(2)_{A}\otimes SU(2)_{B}$. There is a well known isomorphism between classical algebras $\mathfrak{so}(4) \simeq \mathfrak{su}(2) \otimes \mathfrak{su}(2) $ , which allow us to relate the representations of the HLG with those of 
$SU(2)_{A}\otimes SU(2)_{B}$.  The generators of the latter group are related to the rotations, $\mathbf{J}$, and boosts, $\mathbf{K}$, generators as 
\begin{equation}
\mathbf{A}=\frac{1}{2}(\mathbf{J}-i\mathbf{K}),\qquad\mathbf{B}=\frac{1}%
{2}(\mathbf{J}+i\mathbf{K}). \label{AB}%
\end{equation}
The generators of the HLG, $\{\mathbf{J},\mathbf{K}\}$, transform as the components of a second rank antisymmetric tensor $M\indices{_\mu_\nu}$ according to
\begin{equation}
M\indices{^0^i}=K_{i},\qquad M\indices{^i^j}=\epsilon\indices{_i_j_k}J_{k}%
\end{equation}
and for the simplest representations $(j,0)$ and $(0,j)$, here denoted by ``right'' and ``left'' respectively, 
this tensor has the following components
\begin{align}
M_{R}^{0i}  &=\left(  K_{R}\right)_{i},\qquad M_{R}^{ij}=\epsilon_{ijk}\left(  J_{R}\right)_{k},\\
M_{L}^{0i}  &=\left(  K_{L}\right)_{i},\qquad M_{L}^{ij}=\epsilon_{ijk}\left(  J_{L}\right)_{k},
\end{align}
which satisfy%
\begin{equation}
\mathbf{K}_{R}    =i\mathbf{J}_{R}, \qquad \mathbf{K}_{L}   =-i\mathbf{J}_{L}
\label{GenRL}
\end{equation}
where $\mathbf{J}_{R}=\mathbf{J}_{L}=\boldsymbol\tau$ are the conventional $\left(2j+1\right) \times\left(2j+1\right)$ rotation matrices 
for spin $j$ in the $\{\ket{j,m}\}$ basis. Due to these relations each of the second rank antisymmetric tensors $M_{R,L}^{\mu\nu}$,  has 
only $3$ independent components. This lack of independence can be covariantly written as 
\begin{equation}
\widetilde{M}_{R}^{\mu\nu} = -i M_{R}^{\mu\nu}, \qquad 
\widetilde{M}_{L}^{\mu\nu} =  i M_{L}^{\mu\nu}, 
\end{equation}
with the dual tensor defined by
\begin{equation}
\widetilde{M}\indices{^\mu^\nu}\equiv\frac{1}{2}\varepsilon\indices{^\mu^\nu^\alpha^\beta}M_{\alpha\beta},
\end{equation}
where we use the convention $\varepsilon^{0123}=1$.

Now, the generators for the $(j,0)\oplus(0,j)$ representation in this basis (denoted as chiral basis in the following) are
\begin{equation}
M\indices{^\mu^\nu}=\left(
\begin{array}
[c]{cc}%
M_{R}^{\mu\nu} & 0\\
0 & M_{L}^{\mu\nu}
\end{array}
\right),
\end{equation}
and the dual to this tensor is%
\begin{equation}
\widetilde{M}\indices{^\mu^\nu}=\left(
\begin{array}
[c]{cc}%
\widetilde{M}_{R}^{\mu\nu} & 0\\
0 & \widetilde{M}_{L}^{\mu\nu} 
\end{array}
\right) = -i\left(
\begin{array}
[c]{cc}%
M_{R}^{\mu\nu} & 0\\
0 & -M_{L}^{\mu \nu}%
\end{array}
\right)  =-i\chi M\indices{^\mu^\nu}, 
\label{chiten}%
\end{equation}
where $\chi$ is the \textit{chirality} operator%
\begin{equation}
\chi=\left(
\begin{array}
[c]{cc}%
1 & 0\\
0 & -1
\end{array}
\right)  . 
\label{chi}%
\end{equation}
In vector language, for $(j,0)\oplus(0,j)$ we have%
\begin{equation}
\mathbf{J}=\left( 
\begin{array}[c]{cc}%
\mathbf{\boldsymbol\tau} & 0\\
0 & \mathbf{\boldsymbol\tau}%
\end{array}
\right)  ,\qquad
\mathbf{K}=\left(
\begin{array}[c]{cc}%
i\mathbf{\boldsymbol\tau} & 0\\
0 & -i\mathbf{\boldsymbol\tau}%
\end{array}
\right)  .
\label{JKtau}
\end{equation}
These operators are related as%
\begin{equation}
\mathbf{K}=i\chi\mathbf{J}. 
\label{KJ}%
\end{equation}

Parity exchanges $(j,0)\leftrightarrow(0,j)$. Hence, modulo a global phase, in the rest frame the parity operator in the chiral basis reads
\begin{equation}
\Pi=\left( \begin{array}
[c]{cc}%
0 & 1\\ 
1 & 0
\end{array} \right)  . \label{pi}%
\end{equation}
The generators satisfy the conventional Lie algebra%
\begin{equation}
\lbrack M\indices{^\mu^\nu},M\indices{^\alpha^\beta}]=-i\left(  g\indices{^\mu^\alpha}M\indices{^\nu^\beta} - g\indices{^\mu^\beta}M\indices{^\nu^\alpha}-g\indices{^\nu^\alpha}M\indices{^\mu^\beta} + g\indices{^\nu^\beta}M\indices{^\mu^\alpha}\right)  . 
\label{Liealg}%
\end{equation}
It is interesting that for the $(j,0)\oplus(0,j)$ representations the chirality operator can be written as
\begin{equation}
\chi=\frac{i}{4j(j+1)}\widetilde{M}\indices{^\mu^\nu}M\indices{_\mu_\nu}, \label{chicov}%
\end{equation}
i.e. it is proportional to the second Casimir operator of the Lorentz group. As such, it commutes with the generators,%
\begin{equation}
\lbrack\chi,M\indices{^\mu^\nu}]=0, 
\label{chim}
\end{equation}
and a straightforward calculation yields%
\begin{equation}
\{\chi,\Pi\}=0. 
\label{chipi}%
\end{equation}

Notice that Eqs. (\ref{GenRL}) yields an explicit matrix representation for the boost generators of the $(j,0)$ and $(0,j)$ 
representations, which allows us to explicitly construct the corresponding boost operators. Acting with this operator 
on the rest frame states we construct the explicit form of the corresponding states in an arbitrary frame. By extension 
of the $j=1/2$ case, we will call also ``{\it spinors}'' the matrix representation of the states for arbitrary $j$ (chiral 
spinors in this case, see below) even though they do not furnish a representation of a Clifford algebra. Instead, 
as shown in \cite{Gomez-Avila:2013qaa}, they possess of a more complicated algebraic structure.    

The $(j,0)\oplus(0,j)$ representation has $2(2j+1)$ complex degrees of freedom, twice the needed to describe 
a spin $j$ particle, but in contrast with $(j,0)$ and $(0,j)$, it admits a representation for discrete symmetries. 
In other words, these are the irreducible representations for the complete HLG, not just the connected part.
 
The construction of spinors transforming in the $(j,0)\oplus(0,j)$ representation can be done following the same procedure but 
now using the second of Eqs. (\ref{JKtau}). In this case the boost operator is the same for all states in the $(j,0)\oplus(0,j)$ 
and the transformation properties under discrete symmetries of the corresponding spinors are dictated by the specific choice of 
the rest frame spinors. Since parity historically played an important role in the description of electromagnetic and strong 
interactions, states with well defined parity have been considered in the past and the corresponding states in momentum space 
have been constructed using this procedure without reference to a Lagrangian or to an equation of motion for 
$j=1/2, 1, 3/2, 2$ \cite{Ahluwalia:1999ny, DelgadoAcosta:2012yc}. It is worth to remark, however, that spinors with different 
transformation properties under discrete symmetries can also be constructed this way, and specially it is possible to obtain 
explicit representations for Weyl spinors - representations of the Weyl states embedded in this larger space- or Majorana spinors 
if we have an appropriate construction of charge conjugation in the $(j,0)\oplus (0,j)$ representation space. 

In a quantum field theory, the states are the one-particle amplitudes of the corresponding field. This amplitude -the wave function- 
satisfies some equation of motion. We take in this paper the position that this equation of motion is related to the chosen properties 
of the particle in the rest frame, and therefore these properties dictate the structure of the corresponding quantum field theory 
when a suitable Lagrangian is constructed and the gauge principle is used. We study here the case of particles with well defined parity, 
and thus we impose the parity projection condition on the rest frame spinors. In an arbitrary frame this constraint eliminates the 
redundant degrees of freedom, leaving only the $2j+1$ degrees of freedom necessary to describe a spin $j$ and mass $m$ particle, with 
the additional property of well defined parity. The proper -covariant- implementation of the parity projection in an arbitrary frame 
requires to work out the transformation properties under Lorentz transformations of operators acting on the $(j,0)\oplus(0,j)$ 
representation, specially of parity, and to find a basis with definite transformation properties under the Lorentz group. This 
calculation is nontrivial and the details are given in a separate publication \cite{Gomez-Avila:2013qaa}. In that work, an algorithm 
is given for the calculation of the parity-based covariant basis for arbitrary $j$ and the $j=1/2, 1 , 3/2$ have been worked out 
explicitly. We refer the reader to Ref. \cite{Gomez-Avila:2013qaa} for the details, but in order to make the paper as self-contained 
as possible and settle down our notation for the parity-based formalism in the case $j=1$ to be worked out in detail below, we briefly 
review the main results in the next section where we re-derivate the explicit form of one of the the symmetric traceless tensor 
entering this basis for $j=1$ in a way convenient for the purposes of this paper.

\section{Covariant basis for operators in \texorpdfstring{$(j,0)\oplus(0,j)$} constructed from parity.}

The construction of a quantum field theory for particles with well defined parity transforming in the $(j,0)\oplus(0,j)$ 
representation requires that we elucidate the covariant properties of parity and find a basis with well defined properties 
under Lorentz transformations for this space. In the following we will refer to this basis as the covariant basis. 
The $(j,0)\oplus(0,j)$ space has dimension $2(2j+1)$ and there are $4(2j+1)^{2}$ independent matrices acting in this space. 
Any operator $\cal{O}$ acting on this space can be written as a sum of the 
external product of the states%
\begin{equation}
{\cal O}=\sum_{n}c_{n}\Ket{n}\Bra{n} 
\end{equation}
where $\{|n\rangle\}$ is a basis of the representation space $(j,0)\oplus(0,j)$. Using the chiral basis, 
$\{\ket{j,m}_R,\ket{j,m}_L \}$,  this external product space (space of squared matrices of dimension $(2(2j+1))^{2}$) has the symbolic decomposition
\begin{equation}\begin{split}
\left[  (j,0)\oplus(0,j)\right]  \vee \left[  (j,0)\oplus(0,j)\right]   &
=\bigoplus_{i=0}^{2j}[(i,0)\oplus(0,i)]\oplus2(j,j)\\
&  =\left(  [(0,0)\oplus(0,0)]\oplus\lbrack(1,0)\oplus(0,1)]\ \ldots \oplus\lbrack(2j,0)\oplus(0,2j)]\right)  \oplus2(j,j), 
\label{Dec}
\end{split}\end{equation}
where the right hand side enumerates the Lorentz representations under which the operators transform. For every $j$, 
it is possible to construct a set of operators transforming in these Lorentz representations which form a basis 
of the operators acting on $(j,0)\oplus(0,j)$. In general this set contains two Lorentz scalar operators, i.e. 
operators commuting with the Lorentz generators $M\indices{^\mu^\nu}$, which are easily identified as the unit matrix of 
dimension $2(2j+1)$ and the chirality operator $\chi$ in Eq. (\ref{chi}). The decomposition in Eq. (\ref{Dec}) also 
contains six operators transforming in the $(1,0)\oplus(0,1)$ forming a rank-2 anti-symmetric tensor which we identify 
with the generators of the HLG, $M\indices{_\mu_\nu}$. We have also a pair of symmetric traceless tensors transforming in the $(j,j)$. 
It was shown in \cite{Gomez-Avila:2013qaa} that the rest frame parity operator is the time  component of one of these 
symmetric traceless tensors denoted $S^{\mu_1\mu_2...\mu_{2j}}$, and that the second tensor is simply given by 
$\chi S^{\mu_1\mu_2...\mu_{2j}}$. The number of elements in the basis increases with $j$, the remaining operators 
must be explicitly constructed for every $j$ and an algorithm for this construction is given in 
\cite{Gomez-Avila:2013qaa}. 

The explicit matrix form of the operators in the covariant basis for $j=1/2, 1, 3/2$ was given in 
\cite{Gomez-Avila:2013qaa}. In particular, the symmetric traceless tensor to which parity belongs, $S^{\mu_1\mu_2...\mu_{2j}}$, 
of relevance here, was obtained for these values of $j$ by analyzing the transformation properties of the rest frame parity 
operator under Lorentz  transformations in an inductive process. However, once we know that parity is the totally temporal 
part of the symmetric $S$ tensor and we have constructed the boost operator, it is easier to obtain its explicit form just 
boosting the rest frame parity operator since
\begin{equation}
B(p)\Pi B^{-1}(p)=\frac{S^{\mu_1\mu_2...\mu_{2j}}p_{\mu_1}p_{\mu_2}...p_{\mu_{2j}}}{m^{2j}}
\equiv \frac{S_{j}(p)}{m^{2j}}.
\label{parcov}
\end{equation}

In the case $j=1/2$ we have two scalar operators, $1$ and $\chi$, an antisymmetric tensor, $M\indices{_\mu_\nu}$, 
and and two vector operators (the ``symmetric'' operators of rank $2j=1$). 
\begin{equation}
\{1,\chi,S^{\mu},\chi S^{\mu},M\indices{^\mu^\nu}\}, 
\label{basis12}
\end{equation}
where
\begin{equation}
S^{\mu}=\Pi\left(  g^{0\mu}-2iM^{0\mu}\right).
\end{equation}
This is the conventional  set used in the literature up to a $1/2$ factor in $M\indices{_\mu_\nu}$. Boosting the rest frame parity 
operator we get
\begin{equation}
B(p)\Pi B^{-1}(p)=\frac{S^{\mu}p_{\mu}}{m}. 
\label{piboost}%
\end{equation}

In the case $j=1$ there are $36$ independent matrices, two scalars, an antisymmetric tensor, two symmetric traceless tensors and a 
tensor transforming in the $(2,0)\oplus (0,2)$ representation. In this case it can be shown that
\begin{equation}
B(p)\Pi B^{-1}(p)=\frac{S\indices{^\mu^\nu}p_{\mu}p_{\nu}}{m^{2}},
\end{equation}
where $S\indices{^\mu^\nu}$ is a symmetric tensor given by%
\begin{equation}
S\indices{^\mu^\nu}=\Pi\left(  g\indices{^\mu^\nu}-i(g^{0\mu}M^{0\nu}+g^{0\nu}M^{0\mu
})-\{M^{0\mu},M^{0\nu}\}\right)  . \label{st1}%
\end{equation}
This tensor is traceless in the Lorentz indices%
\begin{equation}
S\indices{^\mu_\mu}=0,
\end{equation}
and has $9$ independent components. The second symmetric traceless tensor is given by $\chi S\indices{^\mu^\nu}$.

The tensor transforming in the $(2,0)+(0,2)$ representation is given by
\begin{equation} 
C^{\mu\nu\alpha\beta}=4\{M\indices{^\mu^\nu},M\indices{^\alpha^\beta}\}+2\{M^{\mu\alpha}%
,M^{\nu\beta}\}-2\{M^{\mu\beta},M^{\nu\alpha}\}-8(g^{\mu\alpha}g^{\nu\beta
}-g^{\mu\beta}g^{\nu\alpha}). 
\label{Weyl}%
\end{equation}
It has the symmetries of the Weyl tensor of general relativity, namely%
\begin{equation}
C_{\mu\nu\alpha\beta}=-C_{\nu\mu\alpha\beta}=-C_{\mu\nu\beta\alpha},\quad
C_{\mu\nu\alpha\beta}=C_{\alpha\beta\mu\nu}, \label{Weylsym}%
\end{equation}
the contraction of any pair of indices vanishes and it satisfies the algebraic Bianchi identity%
\begin{equation}
C_{\mu\nu\alpha\beta}+C_{\mu\alpha\beta\nu}+C_{\mu\beta\nu\alpha}=0.
\label{Bianchi}%
\end{equation}
These symmetries leave only $10$ independent components out of the $256$
components of a rank-$4$ tensor.

In summary, in the case of spin $\mathbf{1}$ the basis of matrices with well defined
Lorentz transformation properties is%
\begin{equation}
\{\mathbf{1},\chi,S\indices{^\mu^\nu},\chi S\indices{^\mu^\nu},M\indices{^\mu^\nu},C^{\mu\nu\alpha\beta}\}.
\label{basis1}%
\end{equation}
Important relation for our formalism can be obtained from Eqs. (\ref{chim},\ref{chipi},\ref{st1},\ref{Weyl})%
\begin{equation}
\{\chi,S\indices{^\mu^\nu}\}=0, \qquad [\chi, C^{\mu\nu\alpha\beta}] =0.
\label{chis}%
\end{equation}
In general, for arbitrary $j$ the covariant basis is given by%
\begin{equation}
\{1,\chi,M\indices{^\mu^\nu},S^{\mu_{1}\mu_{2}..\mu_{2j}},\chi S^{\mu_{1}\mu_{2}%
..\mu_{2j}},C^{\mu\nu\alpha\beta},...\}
\end{equation}
where $...$ stands for additional tensor operators transforming in the
$(3,0)\oplus(0,3)\oplus...\oplus(2j,0)\oplus(0,2j)$, whose explicit form can be obtained 
following the algorithm developed in Ref. \cite{Gomez-Avila:2013qaa}.

\section{Parity projectors and free theory.}
The condition for a state transforming in $(j,0)\oplus(0,j)$ to have well defined parity can be written in the rest frame as
\begin{equation}
\mathbb{P}_{\pm}(0)u_{\pm}(0)=u_{\pm}(0),
\end{equation}
where $\mathbb{P}_{\pm}$ stands for the projector over the subspaces of well defined parity in the rest frame given as
\begin{equation}
\mathbb{P}_{\pm}(0)=\frac{1}{2}\left(  1\pm\Pi\right)  . \label{parproj}%
\end{equation}
with the $+(-)$ sign corresponding to the positive (negative) parity case. For an arbitrary  frame, the condition for the state 
to have well defined parity is obtained just boosting this equation. The parity operator can be written in terms of the symmetric
traceless tensor operator transforming in the $(j,j) $ representation according to Eq. (\ref{parcov}), thus the projectors 
in an arbitrary frame  are given by 
\begin{equation}
\mathbb{P}_{\pm}(p)=\frac{1}{2}\left(  1\pm\frac{S_j(p)}{m^{2j}}\right)  . 
\label{parprojm}%
\end{equation}
The operator $S_{j}$ satisfies the following relation
\begin{equation}
\left(  S_{j}(p)\right)  ^{2}=(p^2)^{2j}, \label{S2}%
\end{equation}
which can be used to show that, on-shell, the following relations hold 
\begin{equation}
\left(  \mathbb{P}_{\pm}(p)\right)  ^{2}    =\mathbb{P}_{\pm}(p), ~~~~ \mathbb{P}_{+}(p)\mathbb{P}_{-}(p)=\mathbb{P}_{-}(p)\mathbb{P}_{+}(p)=0.
\label{projprop}
\end{equation}
Considering the projector in Eq. (\ref{parprojm}) the parity projection in an arbitrary frame yields the condition%
\begin{equation}
\left(  \pm S_{j}(p)-m^{2j}\right)  u_{\pm}(p,\lambda)=0. \label{eomp}%
\end{equation}
Writing the corresponding wave function as
\begin{equation}
\psi_{\pm}(x)=u_{\pm}(p,\lambda)e^{-ip\cdot x},
\end{equation}
it is easy to show that it obeys the following equation of motion%
\begin{equation}
\left( \pm S_{j}(i\partial)-m^{2j}\right)  \psi_{\pm}(x)=0. 
\label{eom}%
\end{equation}
In particular, for $j=1/2$ and positive parity this is the conventional Dirac equation
\begin{equation}
\left[iS^\mu\partial_{\mu}-m\right]\psi=0.
\end{equation}

In general, Eqs.(\ref{eom}) can be derived from the following Lagrangians%
\begin{equation}
\mathcal{L}_{\pm}=\overline{\psi_{\pm}}(x)\left( \pm S_{j}(i\partial)-m^{2j}\right)
\psi_{\pm}(x).
\label{lagpm}
\end{equation}
Here, the adjoint spinor is defined as%
\begin{equation}
\overline{\psi}=\psi^{\dagger}\Pi ,
\end{equation}
where $\Pi$ is the rest-frame parity operator in Eq. (\ref{pi}). The $\Pi$ factor here is necessary to make 
a scalar product in $(j,0)\oplus(0,j)$ since a boost operator transforms under $\Pi$ as
\begin{equation}
\Pi B(p)\Pi=B^{-1}(p).
\end{equation}
As will be shown later, in the case of fermions states with opposite parities describe particle-antiparticle system. 
It is then possible to consider just the positive parity case in configuration space and the negative parity is 
incorporated as the antiparticle solution as is done in Dirac theory and the $\pm$ notation in Eq. (\ref{lagpm}) is 
not necessary. However, in the boson case it will turn out that the state with the opposite parity does not correspond 
with the anti-particle thus we get different theories for states with different parities. 

The success of Dirac theory suggest to use the projector in  Eq. (\ref{parprojm}) as a framework to 
incorporate interactions via the gauge principle. There are, however, two subtle points in this proposal. The first 
one concerns the form of Eq. (\ref{parcov}). For every $j$ except for $j=1/2$, a completely antisymmetric tensor can 
be added to the symmetric traceless tensor without modifying the boost operator due to the symmetric $p$ factors. For 
the free theory this is irrelevant, but it becomes crucial for the interacting theory, since the antisymmetric tensor 
generates multipole structures beyond the electric charge. The $j=1/2$ case has only one $p$ factor and does not have 
this freedom, we have fixed multipoles ($e$ and $g=2$ ) only in this case.  

The second subtle point concerns the projector properties in (\ref{projprop}). These relations are satisfied by the 
operators in  Eq. (\ref{parprojm}) only on-shell, hence we denote these operators as the {\it on-shell projectors}. 
These relations cease to be satisfied off-shell. The operators that do satisfy the projector relations off-shell are 
\begin{equation}
\tilde{\mathbb{P}}_{\pm}(p)=\frac{1}{2}\left(  1\pm\frac{S_j(p)}{p^{2j}}\right)  . \label{parprojp}%
\end{equation}
and we denote these operators as the {\it off-shell projectors} in the following.

Considering the use of these projectors as a dynamical principle has the inconvenient that the parity projector 
alone does not fix the mass of the particle. The proper fixing of the mass requires to use the corresponding mass 
projector and the requirement of a local theory -the cancellation of the $p^{2j}$ factors in the denominator- makes 
necessary the use  actually a  power of the mass-projector in the following combination
\begin{equation}
\left(\frac{p^{2}} {m^{2}}\right)^{j} \tilde{\mathbb{P}}_{\pm}(p) u_{\pm}(p,\lambda) 
= \left( \frac{p^{2}}{m^{2}} \right)^{j} \frac{1}{2} \left( 1\pm\frac{S_j(p)}{p^{2j}} \right) u_{\pm}(p,\lambda) 
= u_{\pm}(p,\lambda). 
\label{parmproj}%
\end{equation}

In particular for $j=1/2$ this projection yields the following equation in configuration space
\begin{equation}
[\sqrt{-\partial^2}+ i S^\mu\partial_{\mu}- 2m]\psi(x)=0.
\end{equation}
This formalism  would require to work out the operator $\sqrt{-\partial^2}$. This feature is common to all 
semi-integer spins thus we do not consider this route viable and stick to Eq. (\ref{eom}) in the case of 
fermions. In the case of bosons this problem does not appear and we have both possibilities in the formulation 
of interacting theories.  Below we will study in detail the case $j=1$ and will show that this subtlety is at 
the root of the differences between two of the different formalisms existing in the literature for $j=1$ in 
the $(1,0)\oplus(0,1)$ representation. 

\section{Interacting theory and discrete symmetries.}
In this section we study in detail the discrete symmetries of our formalism, specially charge conjugation. 
In the following we consider in detail positive parity in Eq. (\ref{eom}) but our results in this section are 
also valid for negative parity or the equation derived using the mass and parity projector in Eq. (\ref{parmproj}).

Using the gauge principle we obtain the equation of motion satisfied by $\psi$ interacting with an external 
electromagnetic field as 
\begin{equation}
\left[  S_{j}\left(  i\partial- qA\right)  -m^{2j}\right]  \psi=0,
\end{equation}
where $q$ is the charge of the particle. Taking the complex conjugate of Eq. (\ref{eom}) and multiplying on the 
left by $\eta_{c}\Gamma$ where $\eta_{c}$ is a phase and $\Gamma$ is a matrix in the $(j,0)\oplus(0,j)$ space, we obtain%
\begin{equation}
\left[  (-1)^{2j}\Gamma S_{j}^{\ast}\Gamma^{-1}\left( i \partial+q A\right)-m^{2j}\right]  \psi^{c}=0.
\label{cceom}
\end{equation}
where the conjugate field is given by%
\begin{equation}
\psi^{c}=\eta_{c}\Gamma\psi^{\ast}.
\end{equation}
Requiring $\psi^{c}$ to satisfy the same equation as $\psi$ but with the opposite charge we obtain
\begin{equation}
\Gamma (S^{\mu_{1}\mu_{2}...\mu_{2j}})^{\ast}\Gamma^{-1}=(-1)^{2j}S^{\mu_{1}\mu_{2}...\mu_{2j}}.
\label{stransf}
\end{equation}
The construction of the matrix $\Gamma$ satisfying Eq. (\ref{stransf}) can be
done for general $j$ in terms of time reversal. This is an antilinear operator
and it can always be written as
\begin{equation}
\Theta=U\mathcal{K}, \label{tr}%
\end{equation}
with with $\mathcal{K}$ the complex conjugation operator and $U$ a unitary operator. In general, it is easy 
to show that the action of time reversal on the generators of rotations is given by
\begin{equation}
\Theta\mathbf{J}\Theta^{-1}=U\mathbf{J}^{\ast}U^{-1}=-\mathbf{J,} \label{jtr}%
\end{equation}
and that states with well defined angular momentum transform as
\begin{equation}
\Theta\ket{j, m} = \eta(j)i^{2m}\ket{j, -m}, \label{tjm}%
\end{equation}
with $\eta(j)$ a phase depending only on $j$ (not on $m$). Furthermore, the
squared time reversal operator satisfies%
\begin{equation}
\Theta^{2}=(-1)^{2j}. \label{tr2}%
\end{equation}
We can make a matrix representation for $U$ from Eq. (\ref{tjm}).
Alternatively, if we take the conventional representation of the angular momentum generators where $J_{2}$ 
is purely imaginary and $J_{1,3}$ are real matrices, it is equivalent and easier to make the following representation
\begin{equation}
U=\xi(j)\exp(-i\pi J_{2}), \label{UJ}%
\end{equation}
where $\xi(j)$ is a phase depending only on $j$.

Concerning boosts, it can be shown that the generators transform under time reversal as
\begin{equation}
\Theta(\mathbf{K})\Theta^{-1}=\mathbf{K,} \label{trk}%
\end{equation}
thus the generators in Eq.(\ref{AB}) have the following transformation properties
\begin{align}
\Theta(\mathbf{A})\Theta^{-1}  &  =\frac{1}{2}\Theta(\mathbf{J}-i\mathbf{K}%
)\Theta^{-1}=-\frac{1}{2}\left(  \mathbf{J}-i\mathbf{K}\right)  =-\mathbf{A}%
\label{abtr}\\
\Theta(\mathbf{B})\Theta^{-1}  &  =\frac{1}{2}\Theta(\mathbf{J}+i\mathbf{K}%
)\Theta^{-1}=-\frac{1}{2}\left(  \mathbf{J}+i\mathbf{K}\right)  =-\mathbf{B}%
\nonumber
\end{align}
i.e $\mathbf{A}$ and $\mathbf{B}$ have the same transformation properties as
$\mathbf{J}$.

Using these relations for the specific $(j,0)$ and $(0,j)$ representations, we can see that time reversal 
interchanges these irreps of the HLG. Indeed,
\begin{equation}\begin{split}
\Theta\Lambda_{R}\Theta^{-1}  &  =\Theta\lbrack e^{-i\mathbf{J}\cdot
(\mathbf{\theta}+i\mathbf{\varphi)}}]\Theta^{-1}=e^{i\Theta\mathbf{J}%
\Theta^{-1}\cdot(\mathbf{\theta}-i\mathbf{\varphi)}}=e^{-i\mathbf{J}%
\cdot(\mathbf{\theta}-i\mathbf{\varphi)}}=\Lambda_{L},\label{trlr}\\
\Theta\Lambda_{L}\Theta^{-1}  &  =\Theta\lbrack e^{-i\mathbf{J}\cdot
(\mathbf{\theta}-i\mathbf{\varphi)}}]\Theta^{-1}=e^{i\Theta\mathbf{J}%
\Theta^{-1}\cdot(\mathbf{\theta}+i\mathbf{\varphi)}}=e^{-i\mathbf{J}%
\cdot(\mathbf{\theta}+i\mathbf{\varphi)}}=\Lambda_{R}.
\end{split}\end{equation}
Now, for the $(j,0)\oplus(0,j)$ representation, the construction of charge conjugation can be done in terms 
of $\Theta$ as follows%
\begin{equation}
\mathcal{C}=\left(
\begin{array}
[c]{cc}%
0 & \Theta\\
\Theta^{-1} & 0
\end{array}
\right)  =\left(
\begin{array}
[c]{cc}%
0 & U\\
U^{-1} & 0
\end{array}
\right)  \mathcal{K}\equiv\Gamma\mathcal{K}. \label{cc}%
\end{equation}
where we defined the unitary matrix%
\begin{equation}
\Gamma=\left(
\begin{array}
[c]{cc}%
0 & U\\
U^{-1} & 0
\end{array}
\right)  \label{Gamma}%
\end{equation}
with%
\begin{equation}
U=e^{-i\pi J_{2}}. \label{U}%
\end{equation}
The matrix $U$ satisfies%
\begin{equation}
U^{2}=e^{-i2\pi J_{2}}=(-1)^{2j}=\left\{
\begin{array}
[c]{cc}%
+1 & \text{for bosons}\\
-1 & \text{for fermions}%
\end{array}
\right.  ,
\end{equation}
thus%
\begin{equation}
U^{-1}=\left\{
\begin{array}
[c]{cc}
+U & \text{for bosons}\\
-U & \text{for fermions}%
\end{array}
\right.  ,
\end{equation}
and in both cases%
\begin{equation}
\Gamma^{2}=1.
\end{equation}
Furthermore, since $J_{2}$ is purely imaginary (in the Condon-Shortley phase convention for the $\ket{j, m} $ 
states and in the $\{\ket{j, m}\}$ basis for $(j,0)$ and $(0,j)$), then $U$ is a real matrix. We can check that
\begin{equation}
\Gamma\mathbf{J}^{\ast}\Gamma^{-1} = \left( \begin{array}[c]{cc} 0 & U\\ U^{-1} & 0 \end{array} \right)  
\left(  \begin{array} [c]{cc} \mathbf{J}_{R}^{\ast} & 0\\ 0 & \mathbf{J}_{L}^{\ast} \end{array} \right)  
\left( \begin{array}[c]{cc} 0 & U\\ U^{-1} & 0 \end{array} \right)  
= \left( \begin{array} [c]{cc} U\mathbf{J}_{L}^{\ast}U^{-1} & 0\\ 0 & U^{-1}\mathbf{J}_{R}^{\ast}U \end{array} \right) 
= - \mathbf{J} \end{equation}
The transformation properties of $\mathbf{K}$ are easily obtained from%
\begin{equation}
\mathbf{K}=i\chi\mathbf{J,}%
\end{equation}
and taking into account that%
\begin{equation}
\Gamma\chi^{\ast}\Gamma^{-1}=-\chi
\end{equation}
which yields%
\begin{equation}
\{\mathcal{C},\chi\}=0
\label{acchi}
\end{equation}
and%
\begin{equation}
\Gamma\mathbf{K}^{\ast}\Gamma^{-1}=\Gamma\left(  i\chi\mathbf{J}\right)
^{\ast}\Gamma^{-1}=-i\Gamma\left(  \chi\right)  ^{\ast}\Gamma^{-1}%
(-\mathbf{J)}=-i\chi\mathbf{\mathbf{J}=}=-\mathbf{K.}%
\label{Ktransf}
\end{equation}

Summarizing  %
\begin{equation}
\Gamma {M\indices{_\mu_\nu}}^{\ast}\Gamma^{-1}=-M\indices{_\mu_\nu}.
\end{equation}

Now, using Eqs. (\ref{parcov},\ref{Ktransf}) it is easy to show that
\begin{equation}
\Gamma \left(\frac{S_{j}(p)}{m^{2j}}\right)^{\ast}\Gamma^{-1}=B(p)[ \Gamma \Pi^{\ast}\Gamma^{-1}]B^{-1}(p).
\end{equation}
A straightforward calculation yields  
\begin{equation}
\Gamma \Pi^{\ast}\Gamma^{-1}=(-1)^{2j} \Pi ,
\label{Gammapi}
\end{equation}
thus the key transformation property in Eq. (\ref{stransf}) is satisfied and Eq. (\ref{cceom}) indeed describes 
the motion of a particle with the same mass and opposite electric charge. Interestingly, Eq. (\ref{Gammapi}) can 
be rewritten as  
\begin{equation}%
\begin{array}
[c]{cc}%
\lbrack\mathcal{C},\Pi]=0 & \text{for bosons,}\\
\{\mathcal{C},\Pi\}=0 & \text{for fermions.}%
\end{array}
\end{equation}
Thus, for states of well defined parity, the anti-particle (the particle described by the charge conjugated field) 
has the same parity as the particle in the case of bosons and the opposite parity in the case of fermions. In order 
to work out explicitly this result in a generalization of the well known properties of spin 1/2 spinors, let us make 
a transformation to the basis of well defined parity states. As well known, this is done transforming all observables 
and states with the matrix%
\begin{equation}
M=\frac{1}{\sqrt{2}}\left(
\begin{array}
[c]{cc}%
1 & 1\\
1 & -1
\end{array}
\right)  . \label{transfmatrix}%
\end{equation}
In this basis the states are given as%
\begin{equation}
\psi_{\pi}=\left(
\begin{array}
[c]{c}%
\frac{1}{\sqrt{2}}\left(  \phi_{R}+\phi_{L}\right) \\
\frac{1}{\sqrt{2}}\left(  \phi_{R}-\phi_{L}\right)
\end{array}
\right)  \equiv\left(
\begin{array}
[c]{c}%
\phi_{u}\\
\phi_{v}%
\end{array}
\right)  .
\end{equation}
The Lorentz transformations in this basis are (we use in the following the
suffix $\chi$ for the chiral representation above)%
\begin{equation}
\Lambda_{\pi}=M\Lambda_{\chi}M^{-1}=\frac{1}{2}\left(
\begin{array}
[c]{cc}%
\Lambda_{R}+\Lambda_{L} & \Lambda_{R}-\Lambda_{L}\\
\Lambda_{R}-\Lambda_{L} & \Lambda_{R}+\Lambda_{L}%
\end{array}
\right)  .
\end{equation}
It \ is easy to extract the generators which in this basis read%
\begin{equation}
\mathbf{J}_{\pi}=M\mathbf{J}_{\chi}M^{-1}=\left(
\begin{array}
[c]{cc}%
\boldsymbol\tau & 0\\
0 & \boldsymbol\tau%
\end{array}
\right)  ,\qquad\mathbf{K}_{\pi}=M\mathbf{K}_{\chi}M^{-1}=\left(
\begin{array}
[c]{cc}%
0 & i\boldsymbol\tau\\
i\boldsymbol\tau & 0
\end{array}
\right)  ,
\end{equation}
and similarly for parity and chirality operators%
\begin{equation}
\Pi_{\pi}=M\Pi_{\chi}M^{-1}=\left( \begin{array} [c]{cc} 1 & 0\\ 0 & -1 \end{array} \right)  ,\qquad\chi_{\pi}
=M\chi_{\chi}M^{-1}
=\left( \begin{array} [c]{cc} 0 & 1\\ 1 & 0 \end{array} \right)  .
\end{equation}
Now, the charge conjugation matrix $\Gamma$ in this basis reads
\begin{equation}
\Gamma_{\pi}=M\Gamma_{\chi}M^{-1}=\frac{1}{2}\left(
\begin{array}
[c]{cc}%
U+U^{-1} & -U+U^{-1}\\
U-U^{-1} & -U-U^{-1}%
\end{array}
\right)  =\left\{
\begin{array}
[c]{cc}%
\left(
\begin{array}
[c]{cc}%
U & 0\\
0 & -U
\end{array}
\right)  & \text{for bosons}\\
\left(
\begin{array}
[c]{cc}%
0 & -U\\
U & 0
\end{array}
\right)  & \text{for fermions}%
\end{array}
\right.
\end{equation}
The charge conjugated states in this basis are%
\begin{equation}
\psi_{\pi}^{c}=\Gamma_{\pi}\left(
\begin{array}
[c]{c}%
\phi_{u}^{\ast}\\
\phi_{v}^{\ast}%
\end{array}
\right)  =\left\{
\begin{array}
[c]{cc}%
\left(
\begin{array}
[c]{c}%
U\phi_{u}^{\ast}\\
-U\phi_{v}^{\ast}%
\end{array}
\right)  & \text{for bosons}\\
\left(
\begin{array}
[c]{c}%
-U\phi_{v}^{\ast}\\
U\phi_{u}^{\ast}%
\end{array}
\right)  & \text{for fermions}%
\end{array}
\right.  .
\end{equation}
We remark that since charge conjugation is an anti-linear operator at this level, care must be taken in the 
physical interpretation of results when changing the basis. In our case, however, the transformation matrix 
in Eq. (\ref{transfmatrix}) is real and we can proceed as with conventional linear operators. Clearly, in the 
case of fermions, $u$-type spinors (positive parity states) are mapped by charge conjugation onto $v$-type 
spinors (negative parity sates) and vice versa as is usual in the case of spin 1/2. However, for bosons, 
$u$-type spinors are mapped by charge conjugation onto $u$-type spinors and $v$-type spinors are mapped onto 
$v$-type spinors. This is just an explicit confirmation that in the case of bosons charge conjugation does not 
change the parity of the states. Now the question is if the conjugated fields are independent degrees 
of freedom in the case of bosons. In order to check this point let us work out explicitly the spin $1$ case. The 
charge conjugation operator in the parity basis for $(1,0)\oplus(0,1)$ is given by
\begin{equation}
\psi^{c}=\mathcal{C}\psi\equiv\Gamma\mathcal{K}\psi,
\end{equation}
with%
\begin{equation}
\Gamma_{\pi}=\left(
\begin{array}
[c]{cc}%
U & 0\\
0 & -U
\end{array}
\right)
\end{equation}
and
\begin{equation}
U=e^{-i\pi J_{2}}=\left(
\begin{array}
[c]{ccc}%
0 & 0 & -1\\
0 & 1 & 0\\
-1 & 0 & 0
\end{array}
\right).
\end{equation}
 
The conjugate \spinors\ to $u(0,\lambda)$ are%
\begin{align}
u(0,+)^{c}  &  =\left(
\begin{array}
[c]{c}%
U\phi_{u}^{\ast}(0,+)\\
0
\end{array}
\right)  =\left(
\begin{array}
[c]{c}%
-\phi_{u}^{\ast}(0,-)\\
0
\end{array}
\right)  =-u^{\ast}(0,-)\nonumber\\
u(0,0)^{c}  &  =\left(
\begin{array}
[c]{c}%
U\phi_{u}^{\ast}(0,0)\\
0
\end{array}
\right)  =\left(
\begin{array}
[c]{c}%
\phi_{u}^{\ast}(0,0)\\
0
\end{array}
\right)  =u^{\ast}(0,0)\\
u(0,-)^{c}  &  =\left(
\begin{array}
[c]{c}%
U\phi_{u}^{\ast}(0,-)\\
0
\end{array}
\right)  =\left(
\begin{array}
[c]{c}%
-\phi_{u}^{\ast}(0,+)\\
0
\end{array}
\right)  =-u^{\ast}(0,+).\nonumber
\end{align}
The conclusion is that if the fields are real then the conjugated fields are not independent. Only for truly 
complex fields, the conjugate fields can be independent and couple with the opposite charge to the electromagnetic 
field. This is a similar situation to the scalar field where the coupling to the electromagnetic field requires to 
have a complex field and the electric charge is the charge of the corresponding $U(1)$ symmetry. In the fermion 
case the electric charge is also associated with this symmetry but in this case the parity subspaces are not invariant 
under charge conjugation ($\{C,P\}=0$) and we require the whole $(j,0)\oplus(0,j)$ to make a Poincar\'{e} $P,C$ and $T\,$ 
invariant formalism. In the boson case, parity subspaces are charge conjugation subspaces ($[C,P]=0$) and according 
to Eq. (\ref{trlr}) these subspaces are also invariant under time reversal, thus  
\textit{in the boson case it is possible to make a quantum field theory using only the subspaces of well defined parity 
in $(j,0)\oplus(0,j)$}. In general, in the diagonal 
parity representation of operators (Dirac representation of gamma matrices in the case of spin 1/2), rest-frame 
spinors of positive (negative) parity are of the $u$-type ($v$-type) i.e. have vanishing components in the lower (upper) 
part. In an arbitrary frame, these spinors develop components in the lower (upper) part, but these components are 
not dynamical and dictated by the kinematics. The distinctive property of fermions and bosons is 
that charge conjugation associate a state of different parity in the former case and of the same parity in the latter case. 
The fermion case is transparent in the case $j=1/2$ described by the Dirac theory. The explicit case of spin 1 bosons 
will be worked out in the following.

\section{Electrodynamics of spin 1 bosons}
In this section we consider in detail the case of spin $1$ . From our previous discussion on discrete symmetries 
it is clear that we will be able to formulate independent theories for the different parities and on the 
other side we must take care of the subtleties mentioned in Section II. 

\subsection{Theory based on the on-shell parity projector: Theory I.}
We start with the on-shell projectors in Eq. (\ref{parprojm}) but use the freedom to add  an antisymmetric 
part to the symmetric traceless tensor. The general form of the antisymmetric tensor can be obtained as a linear 
combination of the elements of the covariant basis in Eq. (\ref{basis1}) but the only possibility is a multiple of 
the Lorentz generators tensor thus the equations of motion for particles with positive ($+$) or negative ($-$) 
parity read%
\begin{equation}
\left(  \mp\Sigma\indices{^\mu^\nu}\partial_{\mu}\partial_{\nu}-m^{2}\right)
\psi(x)=0, 
\label{eomt1}%
\end{equation}
where
\begin{equation}
\Sigma\indices{_\mu_\nu}=S\indices{_\mu_\nu}-i\kappa M\indices{_\mu_\nu}\label{tensort1}
\end{equation}
with $\kappa$ a free parameter. These equations can be derived from the following Lagrangians%
\begin{equation}
\mathcal{L}_{\pm}=\overline{\psi}(x)\left(  \mp\Sigma\indices{^\mu^\nu}%
\partial_{\mu}\partial_{\nu}-m^{2}\right)  \psi(x).
\end{equation}
These Lagrangians can be recast as%
\begin{equation}
\mathcal{L}_{\pm}=\mp\partial_{\mu}\left[  \overline{\psi}(x)\Sigma%
\indices{^\mu^\nu}\partial_{\nu}\psi(x)\right]  \pm\partial_{\mu}\overline{\psi}%
(x)\Sigma\indices{^\mu^\nu}\partial_{\nu}\psi(x)-m^{2j}\overline{\psi}(x)\psi(x).
\end{equation}
In the following we skip the surface term and use the lagrangians%
\begin{equation}
\mathcal{L}_{\pm}^{0}=\pm\partial_{\mu}\overline{\psi}(x)\Sigma^{\mu\nu
}\partial_{\nu}\psi(x)-m^{2}\overline{\psi}(x)\psi(x),
\end{equation}
which are Hermitian since the $\Sigma\indices{^\mu^\nu}$ tensor satisfy%
\begin{equation}
\overline{\Sigma\indices{^\mu^\nu}}\equiv\Pi\left(  \Sigma\indices{_\mu_\nu}\right)
^{\dagger}\Pi=\Sigma^{\nu\mu}.
\end{equation}

The Lagrangians for interacting fields of positive ($+)$ or negative ($-$)
parity are obtained using the $U(1)_{em}$ gauge principle as%
\begin{equation}
\mathcal{L}_{\pm}=\pm\overline{D_{\mu}\psi}\Sigma\indices{^\mu^\nu}D_{\nu}\psi
-m^{2}\overline{\psi}\psi=\mathcal{L}^{0}+\mathcal{L}^{int}, \label{Lags}%
\end{equation}
where $D^{\mu}=\partial^{\mu}-ieA^{\mu}$ and $-e$ is the charge of the
particle. 
A straightforward calculation yields the following interacting Lagrangian
\begin{equation}
\mathcal{L}_{_{\pm}int}=\pm ie[\overline{\psi}\Sigma\indices{^\mu^\nu}\partial_{\nu
}\psi-(\partial_{\nu}\overline{\psi})\Sigma^{\nu\mu}\psi]A_{\mu} 
+ e^{2}\overline{\psi}\Sigma\indices{^\mu^\nu}\psi A_{\mu}A_{\nu}.
\end{equation}
In the case of positive parity, the electromagnetic current is%
\begin{equation}
J^{I}_{\mu}(x)=\bar{\psi}[\Sigma\indices{_\mu_\nu}\partial^{\nu}\psi-(\partial^{\nu}\overline{\psi})\Sigma_{\nu\mu}]\psi.
\end{equation}
In momentum space this current reads
\begin{equation}
J^{I}_{\mu}(p,p^{\prime})=\bar{u}(p^{\prime},\lambda^{\prime})\left[
S\indices{_\mu_\nu}(p^{\prime}+p)^{\nu}+i\kappa M\indices{_\mu_\nu}(p^{\prime}-p)^{\nu}\right]  u(p,\lambda).
\end{equation}
The electromagnetic  current admits a Gordon-like decomposition. Indeed, the external spinors satisfy 
\begin{equation}
\frac{S(p)}{m^{2}}u(p,\lambda)=u(p,\lambda), \qquad \bar{u}(p^{\prime},\lambda^{\prime}) \frac{S(p^{\prime})}{m^{2}}=\bar{u}(p^{\prime},\lambda^{\prime}),
\end{equation}
thus
\begin{equation}
J^{I}_{\mu}(p,p^{\prime})   =\overline{u}(p^{\prime},\lambda^{\prime})\left[  \frac
{S(p^{\prime})}{m^{2}}S\indices{_\mu_\nu}p^{\prime\nu}+S\indices{_\mu_\nu}p_{\nu}\frac
{S(p)}{m^{2}} +i\kappa M\indices{_\mu_\nu}(p^{\prime}-p)^{\nu}\right]  u(p,\lambda).
\end{equation}
The symmetric traceless tensor $S\indices{^\mu^\nu}$ in Eq. (\ref{st1}) satisfies the following commutation rules%
\begin{align}
\lbrack S\indices{^\mu^\nu},S\indices{^\alpha^\beta}]  &  =-i\left(  g^{\mu\alpha}M^{\nu\beta
}+g^{\nu\alpha}M^{\mu\beta}+g^{\nu\beta}M^{\mu\alpha}+g^{\mu\beta}M^{\nu
\alpha}\right)  ,\nonumber\\
\left\{  S\indices{^\mu^\nu},S\indices{^\alpha^\beta}\right\}   &  = \frac{4}{3}\left(
g^{\mu\alpha}g^{\nu\beta}+g^{\nu\alpha}g^{\mu\beta}-\frac{1}{2}g^{\mu\nu
}g\indices{^\alpha^\beta}\right)  -\frac{1}{6}\left(  C^{\mu\alpha\nu\beta}+C^{\mu
\beta\nu\alpha}\right)  . 
\label{conms}%
\end{align}
which can be used to show that
\begin{align}
S(p^{\prime})S\indices{^\mu^\nu}  p^{\prime}_{\nu} &=
 p^{\prime2}\left( p^{\prime\mu}+i M\indices{^\mu^\nu}p^{\prime}_{\nu}\right)  \\
S\indices{^\mu^\nu}S(p)p_{\nu} &  =
 p^{2}\left( p^{\mu}-i M\indices{^\mu^\nu}p_{\nu}\right)  ,
\end{align}
and using the on-shell condition $p^{\prime2}=p^{2}=m^{2}$ we get%
\begin{equation}
J^{I}_{\mu}(p,p^{\prime})=\bar{u}(p^{\prime},\lambda^{\prime})\left[
(p^{\prime}+p)_{\mu}+i(1+\kappa) M\indices{_\mu_\nu}(p^{\prime}-p)^{\nu}\right]  u(p,\lambda).
\end{equation}
This is the Gordon-like decomposition of this theory from where it is clear that the particle has a gyromagnetic factor 
\begin{equation}
g=1+\kappa,
\end{equation}
and reveals the free parameter $\kappa$ as an ``anomalous'' contribution to the magnetic moment. A similar calculation 
to the one presented in Ref. \cite{DelgadoAcosta:2012yc} shows that the multipoles  of this particle are given by
\begin{equation}
q=e, \qquad \mu=\frac{e g}{m}= \frac{e(1+\kappa)}{m} \qquad Q=\frac{e(g-1)}{m^2}=\frac{e\kappa}{m^2}.
\end{equation}
 
\subsection{Theory based on the mass, spin and parity off-shell projector: Theory II.}
Now we turn to the theories based on the mass and parity off-shell projector in Eq. (\ref{parmproj}). Similarly to 
the previous case  the equations of motion for particles with positive ($+$) or negative ($-$) parity read%
\begin{equation}
\left(  -\tilde{\Sigma_{\pm}}\indices{^\mu^\nu}\partial_{\mu}\partial_{\nu}-m^{2}\right)
\psi(x)=0, 
\label{eomt2}%
\end{equation}
and the Lagrangian can be written as
\begin{equation}
\mathcal{L}_{\pm}^{0}=\partial_{\mu}\bar{\psi}(x)\tilde\Sigma_{\pm}^{\mu\nu
}\partial_{\nu}\psi(x)-m^{2}\bar{\psi}(x)\psi(x),
\end{equation}
where now the tensor is given by
\begin{equation}
\tilde{\Sigma_{\pm}}\indices{^\mu^\nu}=\frac{1}{2}\left( g\indices{^\mu^\nu}\pm S\indices{_\mu_\nu}-i2\rho M\indices{_\mu_\nu}\right)
\label{tensort2}
\end{equation}
with $\rho$ a free parameter. 
Gauging the theory we get the following interacting Lagrangian
\begin{equation}
\mathcal{L}_{_{\pm}int}=ie[\overline{\psi}\tilde{\Sigma_{\pm}}\indices{^\mu^\nu}\partial_{\nu
}\psi-(\partial_{\nu}\overline{\psi})\tilde{\Sigma_{\pm}}\indices{^\mu^\nu}%
\psi]A_{\mu}+e^{2}\overline{\psi}\tilde{\Sigma_{\pm}}\indices{^\mu^\nu}\psi A_{\mu}A_{\nu}.
\end{equation}

The electromagnetic current in momentum space for this theory reads
\begin{equation}
J_{II}^{\mu}(p,p^{\prime})=\frac{1}{2}\overline{u}(p^{\prime},\lambda^{\prime})\left[  \left(
p^{\prime}+p\right)^{\mu}\pm S\indices{^\mu^\nu}(p^{\prime}+p)_{\nu}+i2\rho M\indices{_\mu_\nu}(p^{\prime}-p)_{\nu}\right]  u(p,\lambda).
\end{equation}

A similar calculation to the previous theory yields the following Gordon-like decomposition
\begin{equation}
J_{II}^{\mu}=\overline{u}(p^{\prime},\lambda^{\prime})\left[  \left(
p^{\prime}+p\right)  ^{\mu}+i\left(\frac{1}{2}+\rho\right)M\indices{^\mu^\nu}\left(  p^{\prime}-p\right)
_{\nu}\right]  u(p,\lambda).
\end{equation}
Particles in this theory have a gyromagnetic factor
\begin{equation}
g=\frac{1}{2}+\rho
\end{equation}
thus the free parameter in this case also corresponds to an ``anomalous'' magnetic moment and the gyromagnetic 
factor inherent to the mass and parity projector is $g=1/2$. The electromagnetic multipole moments of this particle 
are given by
\begin{equation}
q=e, \qquad \mu= \frac{e(\frac{1}{2}+\rho)}{m} \qquad Q=\frac{e(-\frac{1}{2}+\rho)}{m^2}.
\end{equation}
 
\section{Mapping to the antisymmetric field and comparison with existing formalisms.}

\subsection{Parity-based formalism in tensor basis.}
Since the description of massive spin 1 particles has a long history a side by side comparison with existing formalisms for 
the description of massive spin $1$ particles transforming in the $(1,0)\oplus (0,1)$ is mandatory. With this aim we translate our 
formalism for spin $1$ matter fields to tensor language. We denote the 
corresponding field as $F\indices{^\alpha^\beta}(x)$ where $\alpha,\beta$ are Lorentz indices associated to the $(1,0)\oplus(0,1)$ 
representation. In the tensor basis, instead of a pair of \spinor\ indices, every operator has two pairs of internal Lorentz 
indices. The explicit form of the operators in the covariant basis is given by
\begin{align}
\label{basistl}\mathbf{1}_{\alpha\beta\gamma\delta}  &  =\frac{1}{2}%
(g_{\alpha\gamma}g_{\beta\delta}-g_{\alpha\delta}g_{\beta\gamma}),\nonumber\\
\chi_{\alpha\beta\gamma\delta}  &  =\frac{i}{2}\varepsilon_{\alpha\beta
\gamma\delta},\nonumber\\
\left(  M\indices{_\mu_\nu}\right)  _{\alpha\beta\gamma\delta}  &  = -i\left(
g_{\mu\gamma}\mathbf{1}_{\alpha\beta\nu\delta}+g_{\mu\delta}\mathbf{1}%
_{\alpha\beta\gamma\nu}-g_{\gamma\nu}\mathbf{1}_{\alpha\beta\mu\delta
}-g_{\delta\nu}\mathbf{1}_{\alpha\beta\gamma\mu}\right) \\
\left(  S\indices{_\mu_\nu}\right)  _{\alpha\beta\gamma\delta}  &  = g_{\mu\nu
}\mathbf{1}_{\alpha\beta\gamma\delta}-g_{\mu\gamma}\mathbf{1}_{\alpha\beta
\nu\delta}-g_{\mu\delta}\mathbf{1}_{\alpha\beta\gamma\nu}-g_{\gamma\nu
}\mathbf{1}_{\alpha\beta\mu\delta}-g_{\delta\nu}\mathbf{1}_{\alpha\beta
\gamma\mu}\nonumber\\
\left(  C_{\mu\nu\rho\sigma}\right)  _{\alpha\beta\tau\delta}  &  = 32 \mathbf{1}_{\alpha\tau\mu\nu} \mathbf{1}_{\delta\beta\rho\sigma}
-32 \mathbf{1}_{\alpha\tau\rho\sigma} \mathbf{1}_{\beta\delta\mu\nu } + 6 g_{\alpha\tau} X_{\beta\delta\rho\sigma\mu\nu} 
+ 6 g_{\beta\delta} X_{\alpha\tau\rho\sigma\mu\nu} + 8 \mathbf{1}_{\alpha\beta\delta\tau} \mathbf{1}_{\rho\sigma\mu\nu} \nonumber\\ 
& +  16\mathbf{1}_{\alpha\beta\rho\sigma} \mathbf{1}_{\delta\tau\mu\nu} + 16\mathbf{1}_{\alpha\beta\mu\nu} \mathbf{1}_{\delta\tau\rho\sigma} 
- 16\mathbf{1}_{\alpha\delta\rho\sigma} \mathbf{1}_{\beta\tau\mu\nu}  
- 16\mathbf{1}_{\alpha\delta\mu\nu} \mathbf{1}_{\beta\tau\rho\sigma}  \nonumber
\end{align}
where
\begin{equation}
X_{\alpha\tau\rho\sigma\mu\nu} = 2 g_{\alpha \rho} \mathbf{1}_{\mu\nu\tau\sigma} 
- 2 g_{\alpha \sigma} \mathbf{1}_{\mu\nu\tau\rho} -2 g_{\alpha \mu} \mathbf{1}_{\nu\tau\rho\sigma}
+2  g_{\alpha \nu} \mathbf{1}_{\mu \tau\rho\sigma}.
\end{equation}
all the structure previously described can be translated into tensor
language using the operators in Eq. (\ref{basistl}). The Lorentz generators in
tensor basis are purely imaginary whereas the symmetric traceless tensor turns
out to be real. It is easy to show that charge conjugation in this case is
simply the complex conjugation operation, thus real fields are self-conjugated
(Majorana fields) and the coupling to an electromagnetic field requires to
complexify this representation, similarly to the case of scalars. Also, the
commutation of charge conjugation with parity is apparent from the real nature
of $S\indices{^\mu^\nu}$ in tensor basis.

Another simplification in tensor basis concerns the action of parity and the 
``bar'' notation. The rest-frame parity operator in tensor basis is given by
\begin{equation}
\Pi_{\alpha\beta\gamma\delta}=\left(  S_{00}\right)  _{\alpha\beta\gamma
\delta}=\mathbf{1}_{\alpha\beta\gamma\delta}-2g_{0\gamma}\mathbf{1}%
_{\alpha\beta0\delta}-2g_{0\delta}\mathbf{1}_{\alpha\beta\gamma0}.
\end{equation}
When acting on the field tensor this operator selects the components of well defined parity
\begin{equation}
F_{\pi}^{a\beta}\equiv\Pi\indices{^\alpha^\beta_\gamma_\delta}F^{\gamma\delta
}=F\indices{^\alpha^\beta}-2\left(  g_{0}^{\alpha}F^{0\beta}-g_{0}^{\beta}F^{0\alpha
}\right)  ,
\end{equation}
thus%
\begin{equation}
F_{\pi}^{0i}=-F^{0i},\qquad F_{\pi}^{ij}=F^{ij}.
\label{parprojtensor}
\end{equation}

Since our operator projects onto well defined parity states and antiparticles
have the same parity there is no need to carry the ``bar'' operation thorough
and we can use directly the sign in the Lagrangian as done usually, e.g. 
the Lagrangian Eq. (\ref{Lags}) in the tensor basis reads%
\begin{equation}
\mathcal{L}_{\pm}=\left(  D_{\mu}F\indices{^\alpha^\beta}\right)  ^{\dagger}\left(
\Sigma\indices{^\mu^\nu}\right)  _{\alpha\beta\gamma\delta}D_{\nu}F^{\gamma
\delta}\mp  m^{2}\left(  F\indices{^\alpha^\beta}\right)  ^{\dagger}F\indices{_\alpha_\beta}.
\label{Lagten}%
\end{equation}

Special attention deserves the chirality operator in Eqs. (\ref{basistl}). In
tensor basis this transformation is related to the concept of a dual tensor as%
\begin{equation}
\left(  \chi F\right)  \indices{^\alpha^\beta}=i\widetilde{F}\indices{^\alpha^\beta}.
\end{equation}
Chirality operator is purely imaginary in tensor basis making transparent
that Eq. (\ref{acchi}) is satisfied. Also, an explicit calculation shows that Eq. (\ref{chis}) holds.

\subsection{Existing formalisms in the light of the parity-based formalism.}

The $(1,0)\oplus(0,1)$ field has been considered in the literature either in 
the spinor language \cite{Joos:1962qq},\cite{Weinberg:1964cn},\cite{Shay:1968iq}, \cite{Hammer:1968zz},\cite{Tucker:1971bi}, 
or in the form of an antisymmetric tensor field \cite{Ecker:1988te}. Concerning the spinor 
formalisms, in Refs. \cite{Joos:1962qq},\cite{Weinberg:1964cn} 
the following equation of motion was put forth%
\begin{equation}
\left[  S\indices{^\mu^\nu}\partial_{\mu}\partial_{\nu}+m^{2}\right]  \psi
_{+}(x)=0,\label{JWp}%
\end{equation}
where $\psi_{+}(x)=u(p,\lambda)e^{-ip\cdot x}$ is the $(1,0)\oplus(0,1)$ wave
function. The corresponding \spinor\ $u(p,\lambda)$ satisfies the condition%
\begin{equation}
\left[  S\indices{^\mu^\nu}p_{\mu}p_{\nu}-m^{2}\right]  u(p,\lambda)=0.
\end{equation}
From the perspective of our construction in Eq. (\ref{eomt1}), this is the projection over subspaces of positive 
parity in $(1,0)\oplus(0,1)$, and corresponds to the particular value $\kappa=0$ of the positive parity case of 
Theory I discussed above (see Eq.(\ref{eomt1})). The corresponding equation for negative parity states differ from 
this equation just by a relative sign between the mass term and the $S(p)$ operator,
\begin{equation}
\left[  -S\indices{^\mu^\nu}\partial_{\mu}\partial_{\nu}+m^{2}\right]  \psi
_{-}(x)=0.\label{JWm}%
\end{equation}
It is worth to remark that the construction of discrete symmetries above, specially charge conjugation, apply to 
this theory and the charge conjugated fields have the same parity as the field itself. There are claims in the 
literature on the possibility that the charge conjugated vector bosons in the $(1,0)\oplus(0,1)$ have the opposite 
parity to the boson itself \cite{Ahluwalia:1993zt}. At odds with our construction, the charge conjugation
operation used in \cite{Ahluwalia:1993zt} is obtained from the interacting theory imposing that the field satisfy 
Eq.(\ref{JWp}) and the conjugate field satisfy a \textit{different} equation, the one satisfied by the field of 
opposite parity in Eq.(\ref{JWm}).

The electrodynamics of spin 1 bosons in the Joos-Weinberg formalism has been discussed in the literature 
\cite{Eeg:1972us},\cite{Eeg:1973xs}, being considered as a phenomenological approach due to the ``unphysical''
solutions inherent to this formalism \cite{Hammer:1968zz}. The problem seems to be that the propagator of the theory, 
in addition to the conventional pole at $p^2=m^2$, has an unphysical pole at $p^{2}=-m^{2}$. However, these conclusions 
are based on a naive calculation of the two-point Green function. The proper calculation of Green functions 
requires to work out the corresponding quantum field theory and in particular the calculation of the physical causal 
propagator needs an appropriate handling of the poles for which the conventional $m^2\to m^2 -i\varepsilon$ prescription
is not enough. This calculation is beyond the scope of this paper and will be published elsewhere. Below we discuss 
the naive calculation from the perspective of the most general lagrangian in the tensor formalism. 

In order to avoid the ``problems'' detected for the massive Joos-Weinberg formalism, new proposals with the mass 
shell condition as an auxiliary condition were studied in \cite{Good:1964} which eventually lead to Shay an Good 
\cite{Shay:1968iq} to propose the following modified equation
\begin{equation}
\left[  \partial^{2}+S\indices{^\mu^\nu}\partial_{\mu}\partial_{\nu}+2m^{2}\right]
\psi(x)=0.
\end{equation}
Performing a non-relativistic expansion of the gauged theory they conclude that it describes particles with 
a gyromagnetic factor $g_{SG}=\frac{1}{2}$. The same equation was considered in\cite{Tucker:1971bi} where the 
electrodynamics of a spin boson is developed in this context. In that work, this equation is the first of a 
set of equations for high spin. A naive calculation of the propagator for these theories yields an unphysical pole at $p^{2}=0$ 
for all $j$ except for $j=1$ which corresponds to the Shay-Good equation. From the perspective of our parity-based 
construction, the Shay-Good equation corresponds to the particular value $\rho=0$ of the positive parity case of 
Theory II (see Eq.(\ref{eomt2})). The classical causality properties of the Shay-Good equation when an ``anomalous'' 
magnetic-dipole term is included was studied in \cite{Prabhakaran:1973pr} which conclude that only in the case 
when the gyromagnetic factor take the value $g=1$, i.e. when $\rho=1/2$, the theory is causal. 

Concerning formalisms in the tensor language, aiming to construct an effective field theory description of the 
interactions of spin 1 hadrons, in Ref. \cite{Ecker:1988te} a tensor formalism for spin 1 was developed. This 
formalism is grounded on general arguments and it is worth to make a careful comparison with our results and 
the existing literature. First, they write the most general second order Lagrangian for the antisymmetric 
tensor field as
\begin{equation}
\mathcal{L}=a\partial^{\mu}F_{\mu\beta}\partial_{\nu}F^{\nu\beta}%
+b\partial^{\mu}F_{a\beta}\partial_{\mu}F^{a\beta}+cF\indices{^\alpha^\beta}%
F\indices{_\alpha_\beta}.\label{EGPRLag}%
\end{equation}
This is indeed the most general second order Lagrangian and can be put in a form resembling our Lagrangian using 
the symmetries of the field
\begin{equation}
\mathcal{L}=\partial^{\mu}F\indices{^\alpha^\beta}\left[  \left(  S\indices{_\mu_\nu^G}\right){}\indices{_\alpha_\beta_\gamma_\delta}\right]  \partial^{\nu}F^{\gamma\delta
}+F\indices{^\alpha^\beta}\left[  c1_{\alpha\beta\gamma\delta}\right]  F^{\gamma\delta
}.
\end{equation}
where the most general symmetric tensor allowed by Lorentz covariance is%
\begin{equation}
\left(  S\indices{_\mu_\nu^G}\right)  _{\alpha\beta\gamma\delta}=bg\indices{_\mu_\nu}%
1_{\alpha\beta\gamma\delta}+\frac{a}{4}\left(  g_{\mu\gamma}1_{\alpha\beta
\nu\delta}+g_{\mu\delta}1_{\alpha\beta\gamma\nu}+g_{\nu\gamma}1_{\alpha
\beta\mu\delta}+g_{\nu\delta}1_{\alpha\beta\gamma\mu}\right)  .\label{geno}%
\end{equation}
In terms of the symmetric traceless tensor in Eq. (\ref{basistl}) and skipping the representation indices we get
\begin{equation}
S\indices{_\mu_\nu^G}=\left(  b+\frac{a}{4}\right)  g\indices{_\mu_\nu}-\frac{a}{4}S\indices{_\mu_\nu}.
\end{equation}
The operator of the most general Lagrangian in Eq. (\ref{EGPRLag}) in momentum space is%
\begin{equation}
O^{G}=\left(  b+\frac{a}{4}\right)  p^{2}-\frac{a}{4}S(p)+c\equiv
Ap^{2}+BS(p)+C.
\end{equation}
In a naive calculation, the propagator is simply the inverse of this operator and can be found as follows. Defining the operator%
\begin{equation}
\widetilde{O}^{G}=\alpha p^{2}+\beta S(p)+\kappa
\end{equation}
we obtain the product%
\begin{equation}
\widetilde{O}^{G}O^{G}=f(p^{2})+\left[  \left(  \beta A+\alpha B\right)
p^{2}+\left(  \beta C+\kappa B\right)  \right]  S\left(  p\right)
,\label{OOtprod}%
\end{equation}
with
\begin{equation}
f(p^{2})=\left(  \alpha A+\beta B\right)  p^{4}+\left(  \alpha C+\kappa
A\right)  p^{2}+\kappa C.
\end{equation}
Imposing to have an operator proportional to $\mathbf{1}$ on the r.h.s. of Eq. (\ref{OOtprod}) requires the 
coefficient of $S(p)$ in Eq. (\ref{OOtprod}) to vanish for all $p^{2}$ thus%
\begin{equation}
\beta A+\alpha B=0,\qquad\beta C+\kappa B=0.\label{Snule}%
\end{equation}
Under this condition we get the inverse of $O^{G}$ as%
\begin{equation}
\left(  O^{G}\right)  ^{-1}=\frac{\widetilde{O}^{G}}{f(p^{2})}.
\end{equation}
The poles of the propagator would be the zeros of $f(p^{2})$ which are given by%
\begin{equation}
f(p^{2})=\left(  \alpha A+\beta B\right)  p^{4}+\left(  \alpha C+\kappa
A\right)  p^{2}+\kappa C=0.
\end{equation}
Multiplying by $A/\alpha$ and using Eqs. (\ref{Snule}) we get the condition
\begin{equation}
\left(  A^{2}-B^{2}\right)  p^{4}+2ACp^{2}+C^{2}=0,
\end{equation}
which yields the poles at%
\begin{equation}
p_{1}^{2}=-\frac{C}{A-B}=-\frac{2c}{a+2b}\equiv M_{1}^{2},\qquad p_{2}%
^{2}=-\frac{C}{A+B}=-\frac{c}{b}\equiv M_{2}^{2}.
\label{poles}
\end{equation}
These are the results for the naive poles quoted in Ref. \cite{Ecker:1988te} and in general we have two of them. There 
are only three possibilities to avoid this naive double pole structure: i) $\ A=-B$ \ ($b=0$), ii) $A=B$ ($a=-2b$), 
iii) $B=0$ ($a=0$) . The first two cases were considered in \cite{Ecker:1988te}. In any of
these cases $f(p^{2})$ is linear in $p^{2}$ and there is only one simple pole located at%
\begin{equation}
p^{2}=-\frac{C}{2A}=M^{2}.
\end{equation}
In both of these cases the naive calculation outlined here yields the correct two-point Green function.
For $A=B$ and removing a global factor $2A$ ( taking $A=1/2)$ yields the following operator%
\begin{equation}
O_{+}^{G}=\frac{1}{2}\left(  p^{2}+S(p)\right)  -M^{2}%
\end{equation}
i.e. we obtain the positive parity projection in Eq. (\ref{parproj}). The inverse operator in this case is given by%
\begin{equation}
\left(  O_{+}^{G}\right)  ^{-1}=\frac{\Delta_{+}(p)}{p^{2}-M^{2}},
\end{equation}
with%
\begin{equation}
\Delta_{+}(p)=\frac{-\frac{1}{2}\left(  p^{2}-S(p)\right)  +M^{2}}{M^{2}}.
\end{equation}
In the case $A=-B$, removing a global factor $2A$ ( taking $A=-1/2)$ the
operator is%
\begin{equation}
O_{-}^{G}=\frac{1}{2}\left(  p^{2}-S(p)\right)  -M^{2},
\end{equation}
which is the positive parity projection in our Eq. (\ref{parproj}). The inverse operator in this case is given by%
\begin{equation}
\left(  O_{-}^{G}\right)  ^{-1}=\frac{\Delta_{-}(p)}{p^{2}-M^{2}}.
\end{equation}
with%
\begin{equation}
\Delta_{-}(p)=\frac{-\frac{1}{2}\left(  p^{2}+S(p)\right)  +M^{2}}{M^{2}}.
\end{equation}
These are precisely the propagators in our Theory II for positive and negative parity respectively. The corresponding 
Lagrangians are
\begin{equation}
\mathcal{L}_{A}=b\left(  \partial^{\mu}F_{a\beta}\partial_{\mu}F^{a\beta
}-2\partial^{\mu}F_{\mu\beta}\partial_{\nu}F^{\nu\beta}-M_{A}^{2}%
F\indices{^\alpha^\beta}F\indices{_\alpha_\beta}\right)  =b\left[  \partial_{\mu}F^{a\beta
}\left(  {\Sigma_{+}}\indices{^\mu^\nu}\right)  _{\alpha\beta\gamma\delta}\partial^{\nu
}F^{\gamma\delta}-M_{A}^{2}F\indices{^\alpha^\beta}F\indices{_\alpha_\beta}\right]  ,
\label{AxialL}
\end{equation}
for the axial case and in the vector case it reads%
\begin{equation}
\mathcal{L}_{V}=a\left(  \partial^{\mu}F_{\mu\beta}\partial_{\nu}F^{\nu\beta
}-\frac{1}{2}M_{V}^{2}F\indices{^\alpha^\beta}F\indices{_\alpha_\beta}\right)  =\frac{a}%
{2}\left[  \partial_{\mu}F^{a\beta}\left(  {\Sigma_{-}}\indices{^\mu^\nu}\right)
_{\alpha\beta\gamma\delta}\partial_{\nu}F^{\gamma\delta}-M_{V}^{2}%
F\indices{^\alpha^\beta}F\indices{_\alpha_\beta}\right]  .
\label{VectorL}
\end{equation}
These Lagrangians coincide with the tensor version of Theory II for the specific value  $\rho=0$. 
In \cite{Ecker:1988te} it is shown that in the first case ($A=B$) the degrees of freedom of the vector part ($F^{ij}$) of the tensor are 
frozen (they are not dynamical) and we are left only with the vector part ($F^{0i}$). Similarly, in the second 
case ($A=B$), the axial components are frozen and we are left with the vector ones. In our formalism this is simply 
a consequence of the projection over subspaces of well defined parity as can be seen from Eq. (\ref{parprojtensor}).

The third possibility to avoid the naive double pole structure is $B=0~(a=0)$. In this case there are no further constrictions 
and the corresponding theory describes a degenerate parity-doublet. This possibility corresponds to the case 
$g=0$ of the Poincar\'{e} projector formalism \cite{Napsuciale:2006wr,DelgadoAcosta:2012yc}  for spin $1$ in the 
$(1,0)\oplus (0,1)$ developed in Ref. \cite{Delgado-Acosta:2013nia}. In that work also the possibility of a theory 
based on the projection over parity eigensubspaces was considered. The parity projector is constructed directly from the 
states which in turn are derived from states in the $(1/2,1/2)$ representation.  In the perspective of the present work, 
the tensor in the so obtained projector is of the form of Eq. (\ref{tensort2}) with $\rho=1/2$. The classical causality properties 
of this theory are also analyzed in \cite{Delgado-Acosta:2013nia} concluding that it is causal. This result is consistent 
with the conclusions of Ref. \cite{Prabhakaran:1973pr}.  

Before ending this section we would like to remark that beyond the analyzed values, the naive calculation of the propagator in 
a second order formalism in general yields two poles. The only possibility for getting two symmetric poles at $p^2=M^2$ and 
$p^{2}=-M^{2}$ from Eqs. (\ref{poles}) is to take $A=0$. This correspond to $b=-\frac{a}{4}$ in whose case $S^{G}$ reduces to 
$S$ modulo a global factor. This case corresponds to $\kappa=0$ in our Theory I (Joos-Weinberg formalism in the case of positive 
parity). Interestingly the massless limit ($c=0$) of this lagrangian has been discussed in the literature in tensor language 
for $a=-1$ (see \cite{Chizhov:2011zz} for a review ). Here, the corresponding action has been shown to be conformally invariant and the fields have been named ``antisymmetric tensor matter fields''. A proper choice of the annihilation and creation operators allows for a 
correct canonical quantization in the massless case and interesting applications to hadron physics and Yukawa interactions in 
physics beyond the standard model and further references can be found in  Ref. \cite{Chizhov:2011zz}. The possibility of an action built 
of an antisymmetric second rank tensor field acting as the ``potential'' of a gauge invariant third rank completely antisymmetric 
third rank tensor field discussed in this work corresponds to $a=-1/2$, $b=1/4$, i.e is the same as the Shay-Good theory  
or the formalism used in $R\chi PT$.      
We remark that a proper calculation of the propagator in the massive case requires to work out the corresponding quantum field theory 
and to find an appropriate prescription for the proper  handling of the naive poles. Otherwise we would have the tachyon propagation 
of the second pole which causes the classical causality problems that have been discussed in the literature. The conformal quantization 
of the massless case of the theory based on the on-shell projection (same as Joos-Weinberg or Chizhov theory) done in \cite{Chizhov:2011zz} 
are interesting guidelines in this concern.

Finally, it is clear that the freedom in the choice of the antisymmetric part of  the space-time tensor appearing when boosting the 
rest-frame parity operator is relevant in the sense that it defines crucial properties of the theory such as the multipole moments. 
Indeed, our construction of the covariant basis shows that these terms can contain only the Lorentz generators tensor $M\indices{_\mu_\nu}$ 
which defines not only the value of the gyromagnetic factor but also all multipole moments. The corresponding term in the lagrangian
is an ``anomalous'' magnetic-dipole interaction which is gauge invariant by itself. Furthermore, the matter fields  have mass dimension one 
and this term is dimension four, thus naively renormalizable, and must be included in the tree level lagrangian.  The lesson is 
that we can either consider the theory with the tensors in Eqs. (\ref{tensort1},\ref{tensort2}), or take  $\kappa=0$ and $\rho=0$ and 
include an anomalous magnetic term in the lagrangian. Even more, the mass dimension one of the matter field allows 
also for dimension four self-interactions, which are also naively renormalizable and must be included in the 
Lagrangian. In the following section we analyze the chiral structure of the theories and the reliability of these 
terms in a chiral theory as the standard model. 

\section{Chiral decomposition.}
The Lagrangian for positive parity in the case of Theory I including all terms  of dimension four reads 
\begin{equation}
\mathcal{L}_{I}=\partial^{\mu}\overline{\psi}\left( S\indices{_\mu_\nu}-i\kappa M\indices{_\mu_\nu} \right) \partial^{\nu}\psi
-m^{2} \overline{\psi}\psi + \mathcal{L}_{self},
\end{equation}
where $\mathcal{L}_{self}$ stands for all self-interaction terms which must be constructed from the following bilinears
\begin{equation}
\overline{\psi}\psi ,\quad  \overline{\psi}\chi\psi, \quad \overline{\psi}S\indices{_\mu_\nu} \psi , \quad
\overline{\psi}\chi S\indices{_\mu_\nu} \psi, \quad \overline{\psi}M\indices{_\mu_\nu} \psi, 
\quad \overline{\psi}C_{\mu\nu\alpha\beta} \psi, \quad \overline{\psi}\chi M\indices{_\mu_\nu} \psi, 
\quad \overline{\psi}\chi C\indices{_\mu_\nu_\alpha_\beta} \psi .   
\end{equation} 
The last two bilinears arises from the contractions of the previous two with the Levi-Civita tensor (contractions with the metric 
tensor vanish) which can be rewritten in terms of the chirality operators using the relations 
\begin{equation}
\widetilde{M}\indices{_\mu_\nu}\equiv \frac{1}{2}\epsilon\indices{_\mu_\nu^\rho^\sigma} M_{\rho\sigma} = -i \chi M\indices{_\mu_\nu}, 
\qquad  \widetilde{C}\indices{_\mu_\nu_\alpha_\beta} \equiv \frac{1}{2}\epsilon\indices{_\mu_\nu^\rho^\sigma} C_{\rho\sigma\alpha\beta}
= -i \chi C\indices{_\mu_\nu_\alpha_\beta}.
\end{equation}
There are ten independent non-vanishing terms that can be built from the products of these bilinears, thus  
\begin{equation}\begin{split}
\mathcal{L}_{self} &= c_{1}\left( \overline{\psi}\psi \right)^{2}+c_{2}\left( \overline{\psi}\chi\psi \right)^{2}
+c_{3}\left( \overline{\psi}S\indices{_\mu_\nu} \psi \right)^{2} + c_{4}\left( \overline{\psi}\chi S\indices{_\mu_\nu} \psi \right)^{2} \\
 & + c_{5}\left( \overline{\psi}M\indices{_\mu_\nu} \psi \right)^{2}+c_{6}\left( \overline{\psi}C\indices{_\mu_\nu_\alpha_\beta} \psi \right)^{2}  
+ c_{7}\left( \overline{\psi}\psi \right)\left( \overline{\psi}\chi\psi \right) + c_{8}\left( \overline{\psi}S\indices{_\mu_\nu} \psi \right) \left( \overline{\psi}\chi S\indices{^\mu^\nu} \psi \right)  \\
& + c_{9}\left( \overline{\psi}M\indices{_\mu_\nu} \psi \right) \left( \overline{\psi}\chi M\indices{^\mu^\nu} \psi \right) 
+c_{10}\left( \overline{\psi}C\indices{_\mu_\nu_\alpha_\beta} \psi \right)\left( \overline{\psi}\chi C\indices{^\mu^\nu^\alpha^\beta} \psi \right).
\end{split}\end{equation}
Now, the chirality operator commutes with all the covariant basis elements except for $S\indices{^\mu^\nu}$ and $\chi S\indices{^\mu^\nu}$,  
for which it anti-commutes. 

Chiral fields are naturally defined in terms of the projectors onto subspaces of well defined chirality
\begin{equation}
\psi_{R}=P_{R}\psi \quad \text{and} \quad \psi_{L}=P_{L}\psi,
\end{equation}
where
\begin{equation}
P_{R}=\frac{1}{2}\left(  1+\chi\right)  ,\qquad  P_{L}=\frac{1}{2}\left(  1-\chi\right)  .
\end{equation}
These operators have the following properties%
\begin{equation}
P_{R}+P_{L}=1,\qquad P_{R}P_{L}=0,\qquad P_{R}^{2}=P_{R},\qquad P_{L}^{2}=P_{L},
\end{equation}
and the commutation relations in Eqs. (\ref{chim},\ref{chis}) yield
\begin{equation}
M\indices{^\mu^\nu}P_{R,L}=P_{R,L}M\indices{^\mu^\nu}, \qquad S\indices{^\mu^\nu}P_{R}=P_{L}S\indices{^\mu^\nu}.
\end{equation}

A chiral transformation of the  field is defined by
\begin{equation}
\psi^{\prime} =U\psi\equiv\exp\left(  i\theta \chi\right)  \psi.
\end{equation}

The chiral transformation of the bilinears, in terms of the chiral fields read
\begin{equation}\begin{split}
\left( \overline{\psi}\psi \right) &\to \left( \overline{\psi}U^2 \psi \right) = e^{2i\theta}\left( \overline{\psi}_L\psi_R \right) + e^{-2i\theta} \left( \overline{\psi}_R\psi_L \right) \\
\left( \overline{\psi}\chi\psi \right) &\to \left( \overline{\psi}\chi U^2 \psi \right )= e^{2i\theta}\left( \overline{\psi}_L\psi_R \right) - e^{-2i\theta} \left( \overline{\psi}_R\psi_L \right)\\
\left( \overline{\psi}S\indices{_\mu_\nu} \psi \right) &\to \left( \overline{\psi}S\indices{_\mu_\nu} \psi \right) = \left( \overline{\psi}_R S\indices{_\mu_\nu} \psi_R \right) + \left( \overline{\psi}_L S\indices{_\mu_\nu} \psi_L \right) \\
\left( \overline{\psi}\chi S\indices{_\mu_\nu} \psi \right) &\to \left( \overline{\psi}\chi S\indices{_\mu_\nu} \psi \right) = \left( \overline{\psi}_R S\indices{_\mu_\nu} \psi_R \right) - \left( 
\overline{\psi}_L S\indices{_\mu_\nu} \psi_L \right) \\
\left( \overline{\psi} M\indices{_\mu_\nu} \psi \right) &\to \left( \overline{\psi}M\indices{_\mu_\nu} U^2\psi \right) =  e^{2i\theta}\left( \overline{\psi}_LM\indices{_\mu_\nu} \psi_R \right) + e^{-2i\theta} \left( \overline{\psi}_R M\indices{_\mu_\nu} \psi_L \right) \\
\left( \overline{\psi}C\indices{_\mu_\nu_\alpha_\beta} \psi \right) &\to  \left( \overline{\psi}C\indices{_\mu_\nu_\alpha_\beta}U^2\psi \right) =  e^{2i\theta}\left( \overline{\psi}_L C\indices{_\mu_\nu_\alpha_\beta} \psi_R \right) + e^{-2i\theta} \left( \overline{\psi}_R C\indices{_\mu_\nu_\alpha_\beta} \psi_L \right) \\
\left( \overline{\psi}\chi M\indices{_\mu_\nu} \psi \right) &\to  \left( \overline{\psi}\chi M\indices{_\mu_\nu} U^2\psi \right) =  e^{2i\theta}\left( \overline{\psi}_LM\indices{_\mu_\nu} \psi_R \right) - e^{-2i\theta} \left( \overline{\psi}_R M\indices{_\mu_\nu} \psi_L \right)  \\
\left( \overline{\psi}\chi C\indices{_\mu_\nu_\alpha_\beta} \psi \right) &\to  \left( \overline{\psi}\chi C\indices{_\mu_\nu_\alpha_\beta}U^2\psi \right) = e^{2i\theta}\left( \overline{\psi}_L C\indices{_\mu_\nu_\alpha_\beta} \psi_R \right) - e^{-2i\theta} \left( \overline{\psi}_R C\indices{_\mu_\nu_\alpha_\beta} \psi_L \right).
\end{split}\end{equation}

From these transformation properties, we see that most of the terms in our lagrangian are not chirally invariant. In particular, mass terms and anomalous magnetic moments are forbidden by chiral symmetry. It is straightforward to show that the chirally 
invariant lagrangian is 
\begin{equation}
\mathcal{L}_{\chi}=\partial^{\mu}\overline{\psi}\left( S\indices{_\mu_\nu}\right) \partial^{\nu}\psi + \mathcal{L}^{\chi}_{self},
\end{equation}
with
\begin{equation}\begin{split}
\mathcal{L}^{\chi}_{self} = & a_{1} [\left( \overline{\psi}\psi \right)^{2}-\left( \overline{\psi}\chi\psi \right)^{2}] 
+ a_{2}\left( \overline{\psi}S\indices{_\mu_\nu} \psi \right)^{2} 
+ a_{3}\left( \overline{\psi}\chi S\indices{_\mu_\nu} \psi \right)^{2} 
+ a_{4}[\left( \overline{\psi}M\indices{_\mu_\nu} \psi \right)^{2} - \left( \overline{\psi}\chi M\indices{_\mu_\nu} \psi \right)^{2}]  \\
& + a_{5}[\left( \overline{\psi}C\indices{_\mu_\nu_\alpha_\beta} \psi \right)^{2} - \left( \overline{\psi}\chi C\indices{_\mu_\nu_\alpha_\beta} \psi \right)^{2}] 
+ a_{6}\left( \overline{\psi}S\indices{_\mu_\nu} \psi \right) \left( \overline{\psi}\chi S\indices{^\mu^\nu} \psi \right)  
\end{split}\end{equation}

The decomposition of the chiral Lagrangian in terms of the chiral field reads
\begin{equation}
\mathcal{L}_{\chi}=\partial^{\mu}\overline{\psi_{R}}S\indices{_\mu_\nu}\partial^{\nu}\psi_{R}
+\partial^{\mu}\overline{\psi_{L}}S\indices{_\mu_\nu}\partial^{\nu}\psi_{L}  + \mathcal{L}^{\chi}_{self} ,
\end{equation}
with
\begin{equation}\begin{split}
\mathcal{L}^{\chi}_{self} = & b_{1}\left( \overline{\psi_R}\psi_L \right)\left( \overline{\psi_L}\psi_R \right)
+ b_{2}\left( \overline{\psi_L}S\indices{_\mu_\nu} \psi_L \right)^{2} 
+ b_{3} \left( \overline{\psi_R}S\indices{_\mu_\nu} \psi_R \right)^{2}
+ b_{4}\left( \overline{\psi_R}M\indices{_\mu_\nu}\psi_L \right)\left( \overline{\psi_L} M\indices{^\mu^\nu}\psi_R \right)  \\
&+ b_{5}\left( \overline{\psi_R}C\indices{_\mu_\nu_\alpha_\beta}\psi_L \right)\left( \overline{\psi_L}C\indices{^\mu^\nu^\alpha^\beta}\psi_R \right)
+ b_{6}\left( \overline{\psi_L}S\indices{_\mu_\nu} \psi_L \right) \left( \overline{\psi_R}S\indices{_\mu_\nu} \psi_R \right) . 
\end{split}\end{equation}

In this form, it is clear that this lagrangian is invariant under the following independent transformations of the chiral fields
\begin{equation}
\psi_{R}^{\prime}  =  \exp\left(  i\alpha_{R}\right)  \psi_{R} \qquad
\psi_{L}^{\prime}  = \exp\left(  i\alpha_{L}\right)  \psi_{L}.
\end{equation}
For this Lagrangian, the decoupling of left and right fields allows them to have different interactions with gauge fields thus 
this is an appropriate formalism to attempt the inclusion of spin 1 matter fields in chiral theories like extensions of the 
standard model. In addition the self-interaction terms produce a richer structure than in the spin one-half case. Similar results are obtained for the negative parity case.

As for theory II, the self-interacting lagrangian is the same but the decomposition of the Lagrangian reads%
\begin{equation}\begin{split}
\mathcal{L}_{II} =& \frac{1}{2}\left(  \partial^{\mu}\overline{\psi_{R}}g\indices{_\mu_\nu}\partial^{\nu}\psi_{L}+\partial^{\mu}\overline{\psi_{R}}S\indices{_\mu_\nu}\partial^{\nu} \psi_{R} + \partial^{\mu}\overline{\psi_{L}} S\indices{_\mu_\nu} \partial^{\nu}\psi_{L}\right) 
\\ &- i e\kappa \partial^{\mu}\overline{\psi_{R}}M\indices{_\mu_\nu}\partial^{\nu}\psi_{L}
-m^{2} \overline{\psi_{R}}\psi_{L} +   R \leftrightarrow L  .
\end{split}\end{equation}
The first term in the kinetic piece couples always the left to the right field thus it is not possible to realize chiral symmetry linearly. 
This formalism is therefore appropriate for (and in the tensor basis has been used in the formulation of ) theories with a chiral symmetry 
realized non-linearly. If confirmed that the scalar boson found at the LHC is the Higgs boson, chiral symmetry must be linearly 
realized and this formalism (Theory II) is not appropriate to be used in possible extensions of the standard  model by  spin 1 matter fields.

\section{Conclusions.}
In this work we use our recent parity-based construction of a covariant basis for operators acting on the 
$(j,0)\oplus(0,j)$ representation of the HLG to propose a formalism for the description of particles of 
spin $j$ and well defined parity, transforming in this representation which by extension of the notation 
of spin $1/2$ in the standard model we call {\it matter} fields. 
We show that for all $j$, except for $j=1/2$, there is a freedom in the writing of the covariant form of parity operator 
which is irrelevant for free theories but which becomes important in the interacting case. Using our covariant basis we 
show that this freedom is related to a magnetic-dipole term in the lagrangian. In addition to this freedom, we show 
that we have the choice of using on-shell or off-shell projectors which yield different nonequivalent interacting theories. 

We construct the operators implementing the discrete symmetries of the theory, specially charge conjugation, and show 
that it commutes with parity in the case of bosons and anti-commutes in the case of fermions as expected. As an explicit 
example of an interacting theory we work out the electrodynamics of spin 1 matter bosons finding two nonequivalent interacting 
theories for a given parity. Using the properties of the covariant basis we perform a Gordon-like decomposition of the corresponding 
electromagnetic current and show that in the formulation based on the on-shell projectors the spin 1 matter boson has a 
gyromagnetic factor $g=1+\kappa$ while in the formalism based on the off-shell projectors it has $g=\frac{1}{2}+\rho$, 
where $\kappa$ and $\rho$ are parameters associated to the above mentioned freedom and turn out to be ``anomalous'' magnetic moments.  
We rewrite our formalisms for spin one matter bosons in the form an antisymmetric tensor field and make a comparison with 
existing formalisms  in the literature, either in spinor or tensor language.

Concerning the spinor formalism, in the case $\kappa=0$ and for positive parity, our theory based on the on-shell projectors 
reproduces an equation proposed by Joos \cite{Joos:1962qq} and Weinberg \cite{Weinberg:1964cn}. In the case $\rho=0$ and for 
positive parity our theory based on the off-shell projectors reproduces an equation proposed by Shay and Good in \cite{Shay:1968iq}. 
For arbitrary $\rho$ this theory reproduces the Shay-Good equation with a magnetic-dipole term which has been shown 
in \cite{Prabhakaran:1973pr} to be classically causal only for $\rho=1/2$.
As for the tensor formalism, we show that the conformally invariant action used by Chizhov  in Ref. \cite{Chizhov:2011zz} corresponds to
$\kappa=0$, $m=0$ and positive parity of our theory based on the on-shell projectors, thus it the same as the massless limit
of  the Joos-Weinberg formalism. In the case $\rho=0$ our theory based on the off-shell projectors recover the tensor 
formalism used in chiral perturbation theory with resonances ($R\chi PT$) \cite{Ecker:1988te}, which is the same as the Shay-Good 
theory just written in tensor language. This theory in tensor language and for the specific case $\rho=1$ coincides with a 
recent proposal based on parity projection where the projector is extracted from an explicit construction of the states, 
in turn induced from states in $(1/2,1/2)$ and which has been proved to  propagate causally in an electromagnetic 
background \cite{Delgado-Acosta:2013nia}, in agreement with results in \cite{Prabhakaran:1973pr}.

Naive power counting admits anomalous magnetic-dipole terms and self-interactions at tree level. We perform a chiral decomposition of these theories and using the properties of the covariant basis we show that chiral symmetry can be realized linearly only for the theory based on the on-shell projectors. Chiral symmetry  forbids not only mass terms but also anomalous magnetic dipole terms and some of the self-interaction terms, leaving only six of them.  We conclude that this is the appropriate framework to attempt the incorporation of spin 1 matter bosons in chiral theories like the standard model. 

\begin{center}
{\bf Acknowledgments}
\end{center}
Work supported by CONACyT under project \# 156618. We thank C.A. Vaquera-Araujo for critical comments on a preliminary version of this work.

\bibliography{highspin}

\begin{thebibliography}{30}
\expandafter\ifx\csname natexlab\endcsname\relax\def\natexlab#1{#1}\fi
\expandafter\ifx\csname bibnamefont\endcsname\relax
  \def\bibnamefont#1{#1}\fi
\expandafter\ifx\csname bibfnamefont\endcsname\relax
  \def\bibfnamefont#1{#1}\fi
\expandafter\ifx\csname citenamefont\endcsname\relax
  \def\citenamefont#1{#1}\fi
\expandafter\ifx\csname url\endcsname\relax
  \def\url#1{\texttt{#1}}\fi
\expandafter\ifx\csname urlprefix\endcsname\relax\def\urlprefix{URL }\fi
\providecommand{\bibinfo}[2]{#2}
\providecommand{\eprint}[2][]{\url{#2}}

\bibitem[{\citenamefont{et. al.}(2012)}]{Aad20121}
\bibinfo{author}{\bibfnamefont{A.~G.} \bibnamefont{et. al.}},
  \bibinfo{journal}{Physics Letters B} \textbf{\bibinfo{volume}{716}},
  \bibinfo{pages}{1 } (\bibinfo{year}{2012}), ISSN \bibinfo{issn}{0370-2693},
  \urlprefix\url{http://www.sciencedirect.com/science/article/pii/S037026931200857X}.

\bibitem[{\citenamefont{et. al}(2012)}]{Chatrchyan201230}
\bibinfo{author}{\bibfnamefont{C.~S.} \bibnamefont{et. al}},
  \bibinfo{journal}{Physics Letters B} \textbf{\bibinfo{volume}{716}},
  \bibinfo{pages}{30 } (\bibinfo{year}{2012}), ISSN \bibinfo{issn}{0370-2693},
  \urlprefix\url{http://www.sciencedirect.com/science/article/pii/S0370269312008581}.

\bibitem[{\citenamefont{Shifman}(2012)}]{Shifman:2012na}
\bibinfo{author}{\bibfnamefont{M.}~\bibnamefont{Shifman}}
  (\bibinfo{year}{2012}), \eprint{1211.0004}.

\bibitem[{\citenamefont{Johnson and Sudarshan}(1961)}]{Johnson:1960vt}
\bibinfo{author}{\bibfnamefont{K.}~\bibnamefont{Johnson}} \bibnamefont{and}
  \bibinfo{author}{\bibfnamefont{E.}~\bibnamefont{Sudarshan}},
  \bibinfo{journal}{Annals Phys.} \textbf{\bibinfo{volume}{13}},
  \bibinfo{pages}{126} (\bibinfo{year}{1961}).

\bibitem[{\citenamefont{Velo and Zwanziger}(1969{\natexlab{a}})}]{Velo:1970ur}
\bibinfo{author}{\bibfnamefont{G.}~\bibnamefont{Velo}} \bibnamefont{and}
  \bibinfo{author}{\bibfnamefont{D.}~\bibnamefont{Zwanziger}},
  \bibinfo{journal}{Phys.Rev.} \textbf{\bibinfo{volume}{188}},
  \bibinfo{pages}{2218} (\bibinfo{year}{1969}{\natexlab{a}}).

\bibitem[{\citenamefont{Velo and Zwanziger}(1969{\natexlab{b}})}]{Velo:1969bt}
\bibinfo{author}{\bibfnamefont{G.}~\bibnamefont{Velo}} \bibnamefont{and}
  \bibinfo{author}{\bibfnamefont{D.}~\bibnamefont{Zwanziger}},
  \bibinfo{journal}{Phys.Rev.} \textbf{\bibinfo{volume}{186}},
  \bibinfo{pages}{1337} (\bibinfo{year}{1969}{\natexlab{b}}).

\bibitem[{\citenamefont{Joos}(1962)}]{Joos:1962qq}
\bibinfo{author}{\bibfnamefont{H.}~\bibnamefont{Joos}},
  \bibinfo{journal}{Fortsch.Phys.} \textbf{\bibinfo{volume}{10}},
  \bibinfo{pages}{65} (\bibinfo{year}{1962}).

\bibitem[{\citenamefont{Weinberg}(1964)}]{Weinberg:1964cn}
\bibinfo{author}{\bibfnamefont{S.}~\bibnamefont{Weinberg}},
  \bibinfo{journal}{Phys.Rev.} \textbf{\bibinfo{volume}{133}},
  \bibinfo{pages}{B1318} (\bibinfo{year}{1964}).

\bibitem[{\citenamefont{Weaver et~al.}(1964)\citenamefont{Weaver, Hammer, and
  Good}}]{Weaver:1964zz}
\bibinfo{author}{\bibfnamefont{D.}~\bibnamefont{Weaver}},
  \bibinfo{author}{\bibfnamefont{C.}~\bibnamefont{Hammer}}, \bibnamefont{and}
  \bibinfo{author}{\bibfnamefont{R.}~\bibnamefont{Good}},
  \bibinfo{journal}{Phys.Rev.} \textbf{\bibinfo{volume}{135}},
  \bibinfo{pages}{B241} (\bibinfo{year}{1964}).

\bibitem[{\citenamefont{Tung}(1967)}]{Tung:1967zz}
\bibinfo{author}{\bibfnamefont{W.-K.} \bibnamefont{Tung}},
  \bibinfo{journal}{Phys.Rev.} \textbf{\bibinfo{volume}{156}},
  \bibinfo{pages}{1385} (\bibinfo{year}{1967}).

\bibitem[{\citenamefont{Shay and Good}(1969)}]{Shay:1968iq}
\bibinfo{author}{\bibfnamefont{D.}~\bibnamefont{Shay}} \bibnamefont{and}
  \bibinfo{author}{\bibfnamefont{J.}~\bibnamefont{Good}, \bibfnamefont{R.H.}},
  \bibinfo{journal}{Phys.Rev.} \textbf{\bibinfo{volume}{179}},
  \bibinfo{pages}{1410} (\bibinfo{year}{1969}).

\bibitem[{\citenamefont{Hammer et~al.}(1968)\citenamefont{Hammer, McDonald, and
  Pursey}}]{Hammer:1968zz}
\bibinfo{author}{\bibfnamefont{C.}~\bibnamefont{Hammer}},
  \bibinfo{author}{\bibfnamefont{S.}~\bibnamefont{McDonald}}, \bibnamefont{and}
  \bibinfo{author}{\bibfnamefont{D.}~\bibnamefont{Pursey}},
  \bibinfo{journal}{Phys.Rev.} \textbf{\bibinfo{volume}{171}},
  \bibinfo{pages}{1349} (\bibinfo{year}{1968}).

\bibitem[{\citenamefont{Seetharaman
  et~al.}(1971{\natexlab{a}})\citenamefont{Seetharaman, Jayaraman, and
  Mathews}}]{Seetharaman:1971nz}
\bibinfo{author}{\bibfnamefont{M.}~\bibnamefont{Seetharaman}},
  \bibinfo{author}{\bibfnamefont{J.}~\bibnamefont{Jayaraman}},
  \bibnamefont{and} \bibinfo{author}{\bibfnamefont{P.}~\bibnamefont{Mathews}},
  \bibinfo{journal}{J.Math.Phys.} \textbf{\bibinfo{volume}{12}},
  \bibinfo{pages}{835} (\bibinfo{year}{1971}{\natexlab{a}}).

\bibitem[{\citenamefont{Seetharaman
  et~al.}(1971{\natexlab{b}})\citenamefont{Seetharaman, Jayaraman, and
  Mathews}}]{Seetharaman:1971rg}
\bibinfo{author}{\bibfnamefont{M.}~\bibnamefont{Seetharaman}},
  \bibinfo{author}{\bibfnamefont{J.}~\bibnamefont{Jayaraman}},
  \bibnamefont{and} \bibinfo{author}{\bibfnamefont{P.~M.}
  \bibnamefont{Mathews}}, \bibinfo{journal}{Journal of Mathematical Physics}
  \textbf{\bibinfo{volume}{12}}, \bibinfo{pages}{1620}
  (\bibinfo{year}{1971}{\natexlab{b}}),
  \urlprefix\url{http://link.aip.org/link/?JMP/12/1620/1}.

\bibitem[{\citenamefont{Tucker and Hammer}(1971)}]{Tucker:1971bi}
\bibinfo{author}{\bibfnamefont{R.}~\bibnamefont{Tucker}} \bibnamefont{and}
  \bibinfo{author}{\bibfnamefont{C.}~\bibnamefont{Hammer}},
  \bibinfo{journal}{Phys.Rev.} \textbf{\bibinfo{volume}{D3}},
  \bibinfo{pages}{2448} (\bibinfo{year}{1971}).

\bibitem[{\citenamefont{Eeg}(1972{\natexlab{a}})}]{Eeg:1972us}
\bibinfo{author}{\bibfnamefont{J.}~\bibnamefont{Eeg}},
  \bibinfo{journal}{Lett.Nuovo Cim.} \textbf{\bibinfo{volume}{4S2}},
  \bibinfo{pages}{223} (\bibinfo{year}{1972}{\natexlab{a}}).

\bibitem[{\citenamefont{Eeg}(1972{\natexlab{b}})}]{Eeg:1973xs}
\bibinfo{author}{\bibfnamefont{J.}~\bibnamefont{Eeg}},
  \bibinfo{journal}{Lett.Nuovo Cim.} \textbf{\bibinfo{volume}{5S2}},
  \bibinfo{pages}{591} (\bibinfo{year}{1972}{\natexlab{b}}).

\bibitem[{\citenamefont{Kirchbach et~al.}(2004)\citenamefont{Kirchbach,
  Compean, and Noriega}}]{Kirchbach:2004qn}
\bibinfo{author}{\bibfnamefont{M.}~\bibnamefont{Kirchbach}},
  \bibinfo{author}{\bibfnamefont{C.}~\bibnamefont{Compean}}, \bibnamefont{and}
  \bibinfo{author}{\bibfnamefont{L.}~\bibnamefont{Noriega}},
  \bibinfo{journal}{Eur.Phys.J.} \textbf{\bibinfo{volume}{A22}},
  \bibinfo{pages}{149} (\bibinfo{year}{2004}), \eprint{hep-ph/0411316}.

\bibitem[{\citenamefont{Delgado-Acosta
  et~al.}(2011)\citenamefont{Delgado-Acosta, Napsuciale, and
  Rodriguez}}]{DelgadoAcosta:2010nx}
\bibinfo{author}{\bibfnamefont{E.}~\bibnamefont{Delgado-Acosta}},
  \bibinfo{author}{\bibfnamefont{M.}~\bibnamefont{Napsuciale}},
  \bibnamefont{and}
  \bibinfo{author}{\bibfnamefont{S.}~\bibnamefont{Rodriguez}},
  \bibinfo{journal}{Phys.Rev.} \textbf{\bibinfo{volume}{D83}},
  \bibinfo{pages}{073001} (\bibinfo{year}{2011}), \eprint{1012.4130}.

\bibitem[{\citenamefont{Weinberg}(1995)}]{Weinberg:1995mt}
\bibinfo{author}{\bibfnamefont{S.}~\bibnamefont{Weinberg}},
  \emph{\bibinfo{title}{The Quantum theory of fields. Vol. 1: Foundations}}
  (\bibinfo{publisher}{Cambridge University Press},
  \bibinfo{address}{Cambridge, UK}, \bibinfo{year}{1995}).

\bibitem[{\citenamefont{Prabhakaran and
  Seetharaman}(1973)}]{Prabhakaran:1973pr}
\bibinfo{author}{\bibfnamefont{J.}~\bibnamefont{Prabhakaran}} \bibnamefont{and}
  \bibinfo{author}{\bibfnamefont{M.}~\bibnamefont{Seetharaman}},
  \bibinfo{journal}{Lett.Nuovo Cim.} \textbf{\bibinfo{volume}{7S2}},
  \bibinfo{pages}{395} (\bibinfo{year}{1973}).

\bibitem[{\citenamefont{Ecker et~al.}(1989)\citenamefont{Ecker, Gasser, Pich,
  and de~Rafael}}]{Ecker:1988te}
\bibinfo{author}{\bibfnamefont{G.}~\bibnamefont{Ecker}},
  \bibinfo{author}{\bibfnamefont{J.}~\bibnamefont{Gasser}},
  \bibinfo{author}{\bibfnamefont{A.}~\bibnamefont{Pich}}, \bibnamefont{and}
  \bibinfo{author}{\bibfnamefont{E.}~\bibnamefont{de~Rafael}},
  \bibinfo{journal}{Nucl.Phys.} \textbf{\bibinfo{volume}{B321}},
  \bibinfo{pages}{311} (\bibinfo{year}{1989}).

\bibitem[{\citenamefont{Napsuciale et~al.}(2006)\citenamefont{Napsuciale,
  Kirchbach, and Rodriguez}}]{Napsuciale:2006wr}
\bibinfo{author}{\bibfnamefont{M.}~\bibnamefont{Napsuciale}},
  \bibinfo{author}{\bibfnamefont{M.}~\bibnamefont{Kirchbach}},
  \bibnamefont{and}
  \bibinfo{author}{\bibfnamefont{S.}~\bibnamefont{Rodriguez}},
  \bibinfo{journal}{Eur. Phys. J.} \textbf{\bibinfo{volume}{A29}},
  \bibinfo{pages}{289} (\bibinfo{year}{2006}), \eprint{hep-ph/0606308}.

\bibitem[{\citenamefont{Delgado-Acosta
  et~al.}(2012)\citenamefont{Delgado-Acosta, Kirchbach, Napsuciale, and
  Rodriguez}}]{DelgadoAcosta:2012yc}
\bibinfo{author}{\bibfnamefont{E.}~\bibnamefont{Delgado-Acosta}},
  \bibinfo{author}{\bibfnamefont{M.}~\bibnamefont{Kirchbach}},
  \bibinfo{author}{\bibfnamefont{M.}~\bibnamefont{Napsuciale}},
  \bibnamefont{and}
  \bibinfo{author}{\bibfnamefont{S.}~\bibnamefont{Rodriguez}},
  \bibinfo{journal}{Phys.Rev.} \textbf{\bibinfo{volume}{D85}},
  \bibinfo{pages}{116006} (\bibinfo{year}{2012}), \eprint{1204.5337}.

\bibitem[{\citenamefont{Delgado-Acosta
  et~al.}(2013)\citenamefont{Delgado-Acosta, Kirchbach, Napsuciale, and
  Rodríguez}}]{Delgado-Acosta:2013nia}
\bibinfo{author}{\bibfnamefont{E.}~\bibnamefont{Delgado-Acosta}},
  \bibinfo{author}{\bibfnamefont{M.}~\bibnamefont{Kirchbach}},
  \bibinfo{author}{\bibfnamefont{M.}~\bibnamefont{Napsuciale}},
  \bibnamefont{and}
  \bibinfo{author}{\bibfnamefont{S.}~\bibnamefont{Rodríguez}}
  (\bibinfo{year}{2013}), \eprint{1303.5511}.

\bibitem[{\citenamefont{Gomez-Avila and
  Napsuciale}(2013)}]{Gomez-Avila:2013qaa}
\bibinfo{author}{\bibfnamefont{S.}~\bibnamefont{Gomez-Avila}} \bibnamefont{and}
  \bibinfo{author}{\bibfnamefont{M.}~\bibnamefont{Napsuciale}}
  (\bibinfo{year}{2013}), \eprint{1307.4711}.

\bibitem[{\citenamefont{Ahluwalia and Ernst}(1993)}]{Ahluwalia:1999ny}
\bibinfo{author}{\bibfnamefont{D.~V.} \bibnamefont{Ahluwalia}}
  \bibnamefont{and} \bibinfo{author}{\bibfnamefont{D.~a.} \bibnamefont{Ernst}},
  \bibinfo{journal}{Int.J.Mod.Phys.} \textbf{\bibinfo{volume}{E2}},
  \bibinfo{pages}{397} (\bibinfo{year}{1993}), \eprint{nucl-th/9905047}.

\bibitem[{\citenamefont{Ahluwalia et~al.}(1993)\citenamefont{Ahluwalia,
  Johnson, and Goldman}}]{Ahluwalia:1993zt}
\bibinfo{author}{\bibfnamefont{D.~V.} \bibnamefont{Ahluwalia}},
  \bibinfo{author}{\bibfnamefont{M.}~\bibnamefont{Johnson}}, \bibnamefont{and}
  \bibinfo{author}{\bibfnamefont{J.~T.~a.} \bibnamefont{Goldman}},
  \bibinfo{journal}{Phys.Lett.} \textbf{\bibinfo{volume}{B316}},
  \bibinfo{pages}{102} (\bibinfo{year}{1993}), \eprint{hep-ph/9304243}.

\bibitem[{\citenamefont{Sankaranarayanan and Good}(1965)}]{Good:1964}
\bibinfo{author}{\bibfnamefont{A.}~\bibnamefont{Sankaranarayanan}}
  \bibnamefont{and} \bibinfo{author}{\bibfnamefont{R.}~\bibnamefont{Good}},
  \bibinfo{journal}{Il Nuovo Cimento Series 10} \textbf{\bibinfo{volume}{36}},
  \bibinfo{pages}{1303} (\bibinfo{year}{1965}), ISSN \bibinfo{issn}{0029-6341},
  \urlprefix\url{http://dx.doi.org/10.1007/BF02750706}.

\bibitem[{\citenamefont{Chizhov}(2011)}]{Chizhov:2011zz}
\bibinfo{author}{\bibfnamefont{M.}~\bibnamefont{Chizhov}},
  \bibinfo{journal}{Phys.Part.Nucl.} \textbf{\bibinfo{volume}{42}},
  \bibinfo{pages}{93} (\bibinfo{year}{2011}).

\end{thebibliography}

\end{document}